\documentclass[twocolumn,showpacs,aps,prl,nofootinbib]{revtex4}
\usepackage{graphicx,dcolumn,amsmath,epsfig,color,colordvi,amsfonts} 
\usepackage{fancyhdr,lineno,multirow,rotating,relsize} 
\RequirePackage{xspace}
\def\babar{\mbox{\slshape B\kern-0.1em{\smaller A}\kern-0.1em
    B\kern-0.1em{\smaller A\kern-0.2em R}}}
\def\CP                {\ensuremath{C\!P}\xspace}
\def\piz   {\ensuremath{\pi^0}\xspace}
\def\Dbar  {\kern 0.2em\overline{\kern -0.2em D}{}\xspace}
\def\Dz    {\ensuremath{D^0}\xspace}
\def\Dzb   {\ensuremath{\Dbar^0}\xspace}
\def\to    {\ensuremath{\rightarrow}\xspace}
\def\invfb   {\ensuremath{\mbox{\,fb}^{-1}}\xspace}
\def\KS    {\ensuremath{K^0_{\scriptscriptstyle S}}\xspace} 
\newcommand{\gevc}{\ensuremath{{\mathrm{\,Ge\kern -0.1em V\!/}c}}\xspace}
\newcommand{\gevcc}{\ensuremath{{\mathrm{\,Ge\kern -0.1em V\!/}c^2}}\xspace}
\newcommand{\mevcc}{\ensuremath{{\mathrm{\,Me\kern -0.1em V\!/}c^2}}\xspace}
\newcommand{\Dzkkpz}{\ensuremath{\Dz \to K^{-}K^{+}\piz}}
\newcommand{\Dzpppz}{\ensuremath{\Dz \to \pi^{-}\pi^{+}\piz}}
\newcommand{\Dzkspp}{\ensuremath{\Dz \to \KS\pi^{+}\pi^{-}}}
\newcommand{\Dskkp}{\ensuremath{D_s^+ \to K^{+}K^{-}\pi^{+}}}
\def\cbar  {\ensuremath{\overline c}\xspace}
\def\ubar  {\ensuremath{\overline u}\xspace}
\def\btou{\ensuremath{b\to u\cbar s}}
\def\btoc{\ensuremath{b\to c\ubar s}}
\def\ppp{\ensuremath{\pi^+\pi^-\pi^0}}

\def\btodkgen{B\to D^{(*)0}K^{(*)}}
\def\ppp{\ensuremath{\pi^+\pi^-\pi^0}}
\def\dtoppp{\ensuremath{D\to \ppp}}
\def\bpmtodkpm{\ensuremath{B^\pm\to DK^\pm}}
\def\decaychain{\ensuremath{B^\pm \to D_{\ppp} K^\pm}}

\long\def\inst#1{\par\nobreak\kern 4pt\nobreak
    {\it #1}\par\vskip 10pt plus 3pt minus 3pt}
\def\figurebox#1#2#3{
    \def\arg{#3}
    \ifx\arg\empty
    {\hfill\vbox{\hsize#2\hrule\hbox to 
	#2{\vrule\hfill\vbox to #1{\hsize#2\vfill}\vrule}\hrule}\hfill}
    \else {\hfill\epsfbox{#3}\hfill}
    \fi}
\begin{document}
\title{{\large \bf Charmed Meson Dalitz Plot Analyses at \babar}}
\author{Kalanand Mishra\footnote{Email:kalanand$@$slac.stanford.edu} 
(for the \babar\ collaboration)}
\affiliation{University of Cincinnati, Cincinnati, Ohio 45221, USA}
\date{\today}
\begin{abstract}
We report recent results of the Dalitz plot analyses of $D$ and 
$D_S$ decays performed by the \babar\ collaboration, and  
point out some of the important applications of these results.  
\end{abstract}
\maketitle
\section{Introduction}
The amplitudes describing $D$ and $D_s$ meson weak decays into 
final states with three pseudo-scalers  
are dominated by intermediate resonances that lead to highly nonuniform 
intensity distributions in the available phase space. The results  
of the Dalitz plot analysis of these decays are playing increasingly 
important role in flavor physics, particularly in the extraction of the 
$C\!P$-violating phase 
$\gamma = \arg{\left(- V^{}_{ud} V_{ub}^\ast/ V^{}_{cd} V_{cb}^\ast\right)}$ 
of the quark mixing (\textit{i.e.,} CKM) matrix by exploiting 
interference structure in the 
$D$ Dalitz plot from the decay $B^{\pm}\to D K^{\pm}$~\cite{myGamma} 
and in the measurement of \Dz--\Dzb\ mixing parameters.
\section{Detector}
We perform these analyses using $e^+e^-$ 
collision data collected at and around 10.58~GeV center-of-mass (CM) energy
with the \babar\ detector~\cite{detector} at the PEP-II storage ring. 
Tracking of charged particles is 
provided by silicon detector and a drift chamber  
operating in a 1.5-T magnetic field. 
Particle types are identified using specific ionization energy 
loss measurements in the two tracking devices and  Cherenkov photons 
detected in a ring-imaging detector. The energy of photons and 
electrons is measured with an electromagnetic calorimeter. In case of 
neutral $D$-meson decays, we distinguish \Dz from \Dzb by reconstructing  
the decays $D^{*+}\to\Dz\pi^{+}$ and $D^{*-}\to\Dzb\pi^{-}$.
For each decay mode, we estimate the signal 
efficiency as a function of position in the Dalitz plot 
using simulated signal events generated uniformly in the 
available phase space, subjected to the same reconstruction 
procedure applied to the data, and corrected for differences 
in particle-identification rates in data and simulation.

\section{Dalitz plot parametrization}
The complex quantum mechanical amplitude $\cal{A}$ that describes 
decays to three particles $A$, $B$ and $C$ in the final state 
can be characterized as a coherent sum of 
all relevant quasi-two-body $D/D_s\to(r\to AB)C$ isobar model resonances, 
${\cal{A}} = \sum_r a_r e^{i\phi_r} A_r(s)$. Here $s=m_{AB}^2$, and 
$A_r$ is the 
resonance amplitude. We obtain the coefficients $a_r$ and $\phi_r$ from 
a likelihood fit. The probability density function for signal events 
is $\left| \cal{A} \right|^2$.

\indent Unless stated otherwise, for \textit{S-}, \textit{P-}, and 
\textit{D-}wave (spin = 0, 1, and 2, respectively) 
resonant states we use the Breit-Wigner amplitude:
\begin{eqnarray}
A_{BW}(s) &=& {{\cal{M}}_L(s,p)}\hspace{2pt}{1\over{M_0^2-s-iM_0\Gamma(s)}},\\
\Gamma(s) &=& \Gamma_0\Big({M_{0}\over \sqrt{s}}
\Big)\Big({p\over{p_{0}}}\Big)^{2L+1} {\Big[{{\cal{F}}_L(p) \over 
      {\cal{F}}_L(p_0)}\Big]^2},
\end{eqnarray}
\noindent where $M_0$ ($\Gamma_0$) is the resonance mass (width)~\cite{pdg}, 
$L$ is the angular momentum quantum number, $p$ is the momentum of either 
daughter in the resonance rest frame, and $p_0$ is the value of $p$ when s = 
$M_0^2$. The function ${\cal{F}}_L$ is the  Blatt-Weisskopf barrier 
factor~\cite{bw}: ${{{\cal{F}}_0}}$ = 1, ${{{\cal{F}}_1}}$ = 
$1/\sqrt{ 1+ {Rp}^2}$, and ${{{\cal{F}}_2}}$ = 
$1/\sqrt{ 9+ 3 {Rp}^2 + {Rp}^4}$, where we take the meson radial parameter $R$ 
to be 1.5 GeV${}^{-1}$. The quantity ${{\cal{M}}_L}$ is 
the spin part of the amplitude: ${{\cal{M}}_0}$ = constant, 
${{\cal{M}}_1} \propto -2 \vec{p_A}.\vec{p_C}$, and 
${{\cal{M}}_2}$ $\propto {4\over 3} \left
[3{(\vec{p_A}.\vec{p_C})}^2 - {|\vec{p_A}|}^2.{|\vec{p_C}|}^2\right]$, where 
$\vec{p_i}$ is the 3-momentum of particle $i$ in the resonance rest frame.
The fit fraction for a resonant 
process $r$ is defined as 
{\footnotesize{ $f_r \equiv \int \left|a_r A_r\right|^2 
d\tau / \int \left|{\cal{A}}\right|^2 d\tau$,}}
where $d\tau$ is a phase-space element. 
Due to interference among the contributing amplitudes, the $f_r$ do 
not sum to one in general.
In all cases, we model small incoherent background empirically from data.

\section{Angular moments}
For $D$ and $D_s$ decays to three spinless particles, the Dalitz plot  
uniquely represents the kinematics of the final state. The angular 
distributions provide further information on the detailed event-density 
variations in various regions of the phase space in a different form. 
We define the helicity angle $\theta_H$ for decays $\Dz\to(r\to AB)C$ as 
the angle between the momentum of $A$ in the $AB$ rest frame and the 
momentum of $AB$ in $D^0$ rest frame. 
The moments of the cosine of the helicity angle, 
$Y_l^0(\cos\theta_H)$, are defined as the efficiency-corrected invariant mass
distributions of events when weighted by spherical harmonic functions 
\begin{equation}
Y_l^0(\theta_H) = \sqrt{1 \over {2\pi}}~ P_l(m), 
\end{equation}
where $m$ is the invariant mass of 
the $AB$ system and the $P_l$ are Legendre polynomials of order $l$:
\begin{equation}
\int\limits_{-1}^1{P_l(x)~ P_n(x)~ dx} =  \delta_{ln}.
\end{equation}

These angular moments have an obvious physical significance. 
Since spherical harmonic functions are the eigen-functions of the angular 
momentum, the Dalitz plot of a three-body decay can be represented by 
the sum of an infinite number of spherical harmonic moments in any 
two-body channel. In a region of the Dalitz plot 
where \textit{S-} and \textit{P-}waves in a single channel dominate, their 
amplitudes are given by the following Legendre polynomial moments,
\begin{eqnarray}
P_0 &=& \frac{{\left| S \right|}^2 + {\left| P \right|}^2}{\sqrt{2}},\nonumber \\ 
P_1 &=& {\sqrt{2}}{\left| S \right|}{\left| P \right|}~\cos\theta_{SP},\nonumber \\
P_2 &=& {\sqrt{2 \over5}} ~{\left | P \right |^2},
\label{eq:pwa}
\end{eqnarray}
\noindent where $\left| S \right|$ and $\left| P \right|$ are, respectively, 
the magnitudes of the \textit{S-} and \textit{P-}wave amplitudes, and 
$\theta_{SP} = \theta_\textit{S} - \theta_P$ is the relative phase between 
them. It is worth noting that this partial-wave analysis is valid, in 
the absence of higher spin states, only if no interference 
occurs from the crossing channels.

\section{Dalitz plot analysis of \Dzkkpz}

\begin{figure}[!htbp]
\begin{tabular}{cc} 
\includegraphics[width=0.235\textwidth]{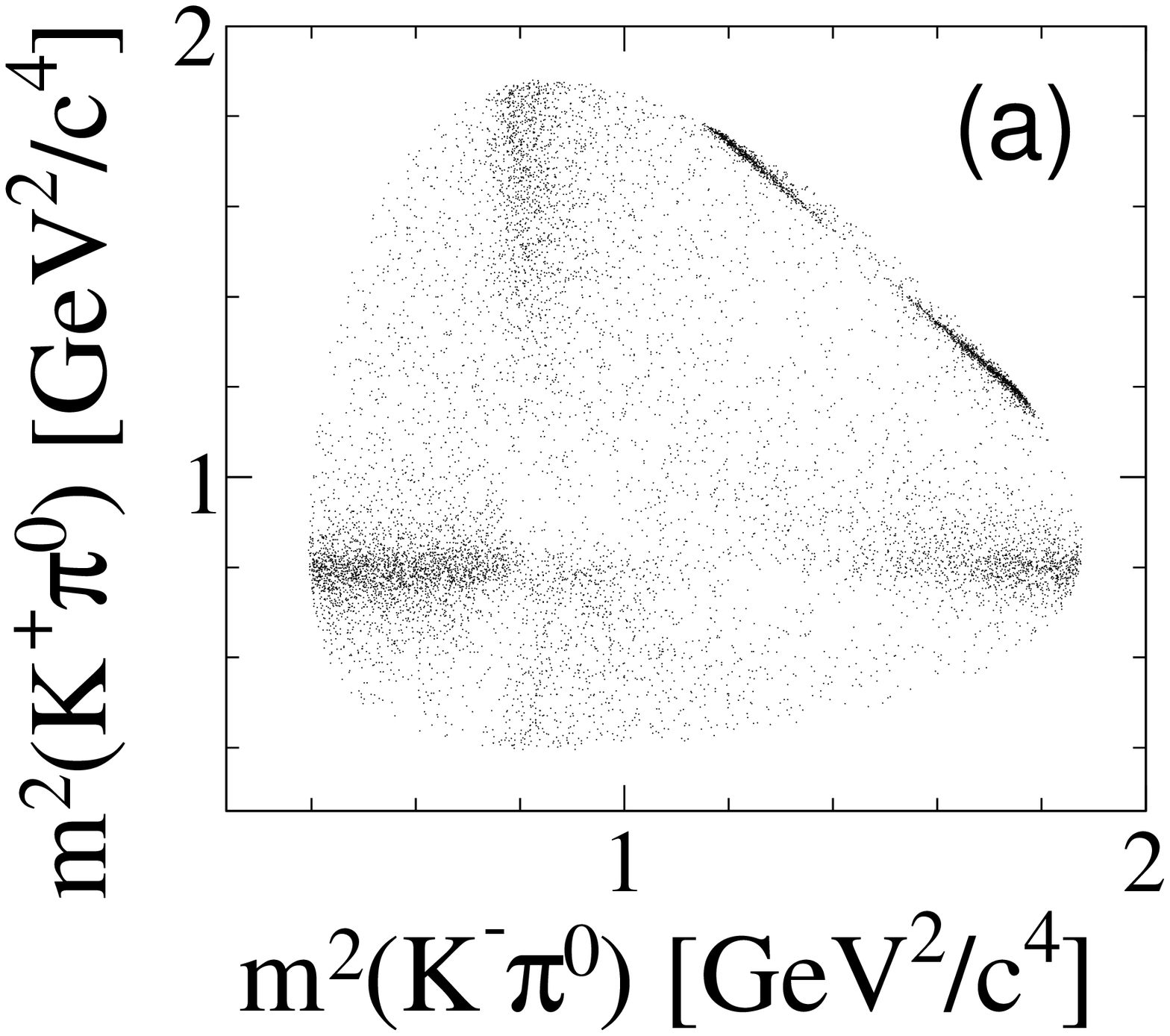}&
\includegraphics[width=0.235\textwidth]{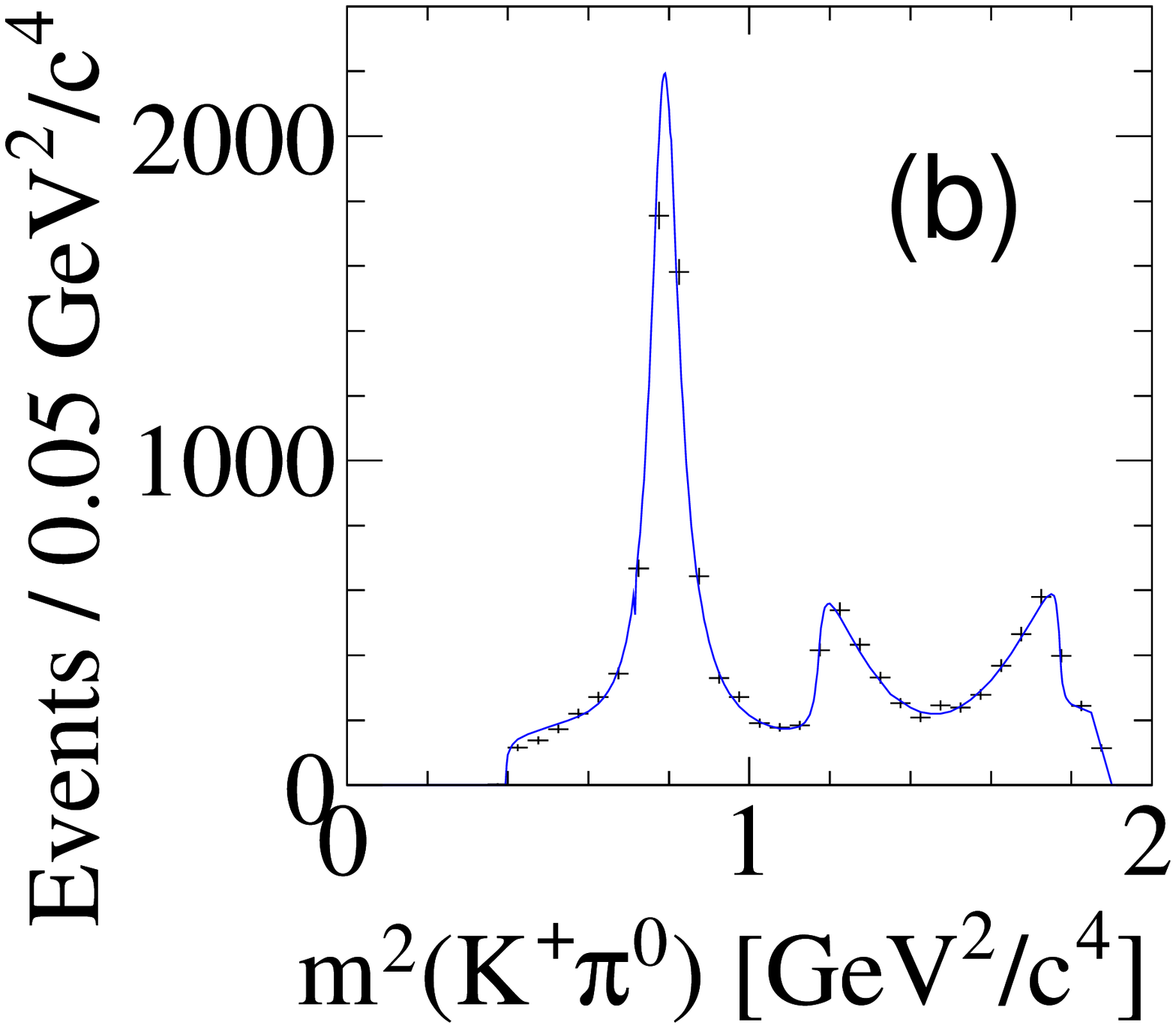}\\
\end{tabular}
\begin{tabular}{cc} 
\includegraphics[width=0.235\textwidth]{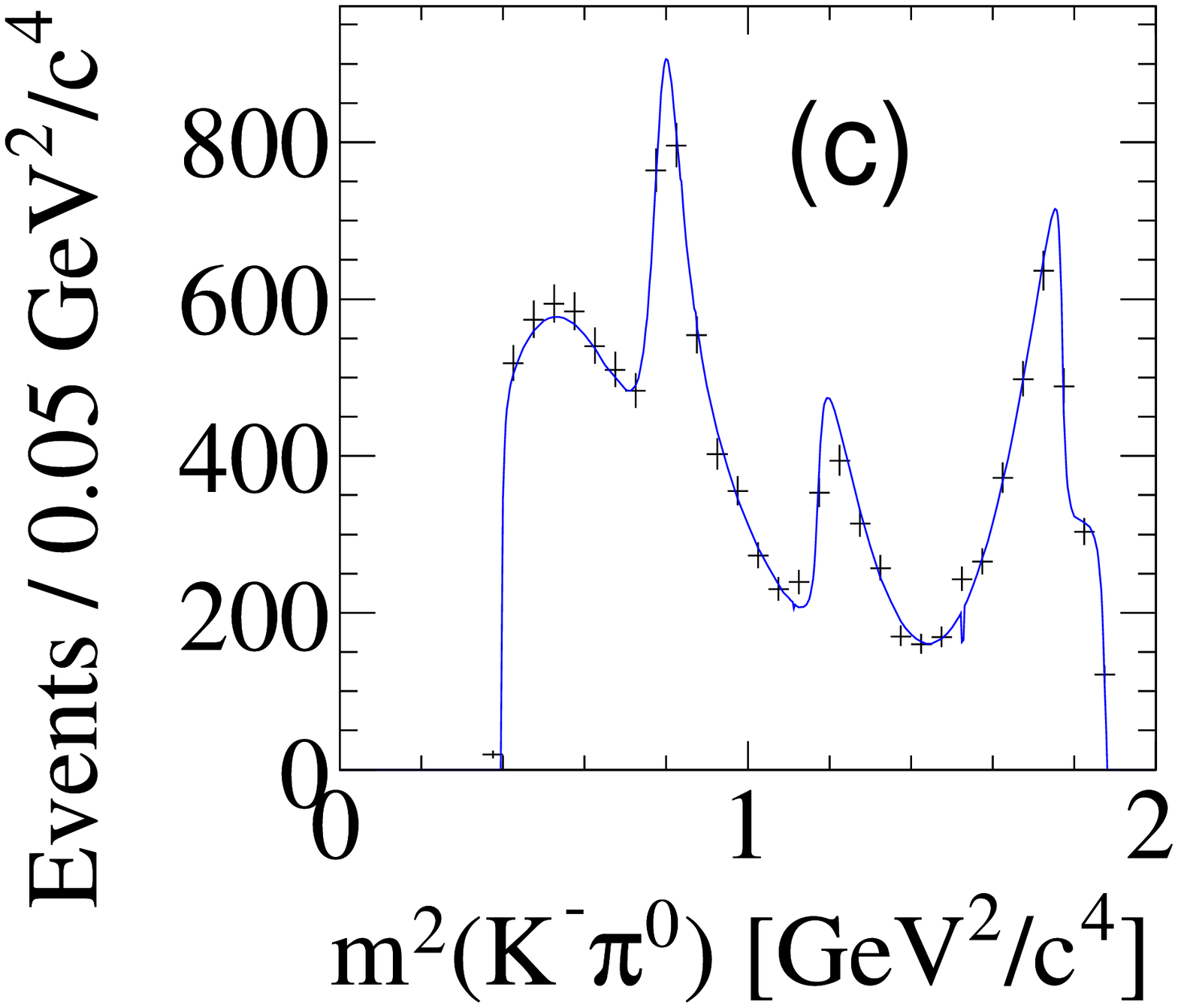}&
\includegraphics[width=0.235\textwidth]{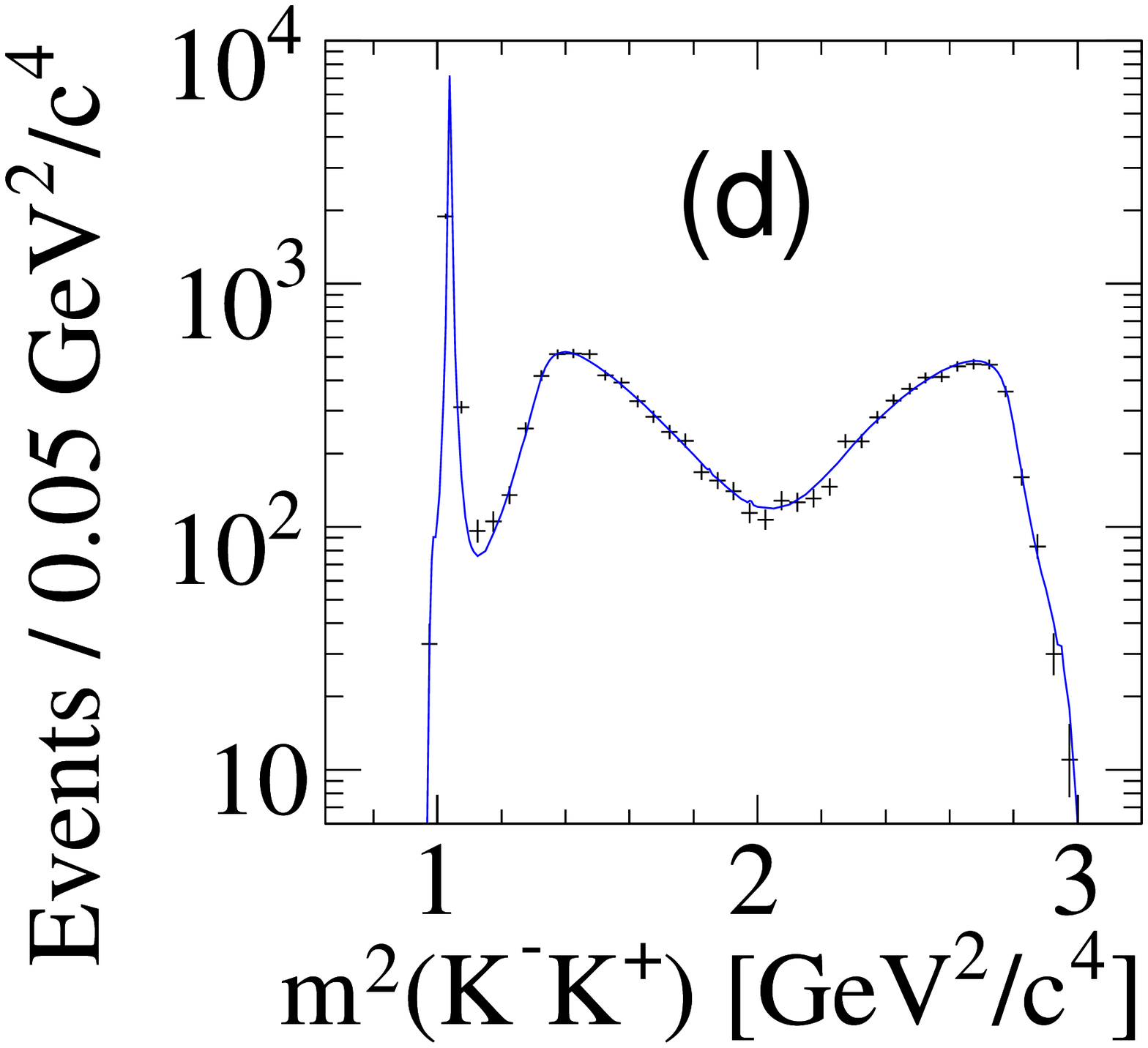}\\
\end{tabular}
\caption{Dalitz plot for $D^0 \to K^- K^+ \piz$~\cite{mykkpi0} 
data (a), and the corresponding squared invariant mass projections (b--d). 
In plots~(b--d), the dots with error bars are data 
points and the solid lines correspond to the best isobar fit models.}
\label{Fig1}
\end{figure}

\begin{figure}[!htbp]
\begin{tabular}{cc} 
\includegraphics[width=0.24\textwidth]{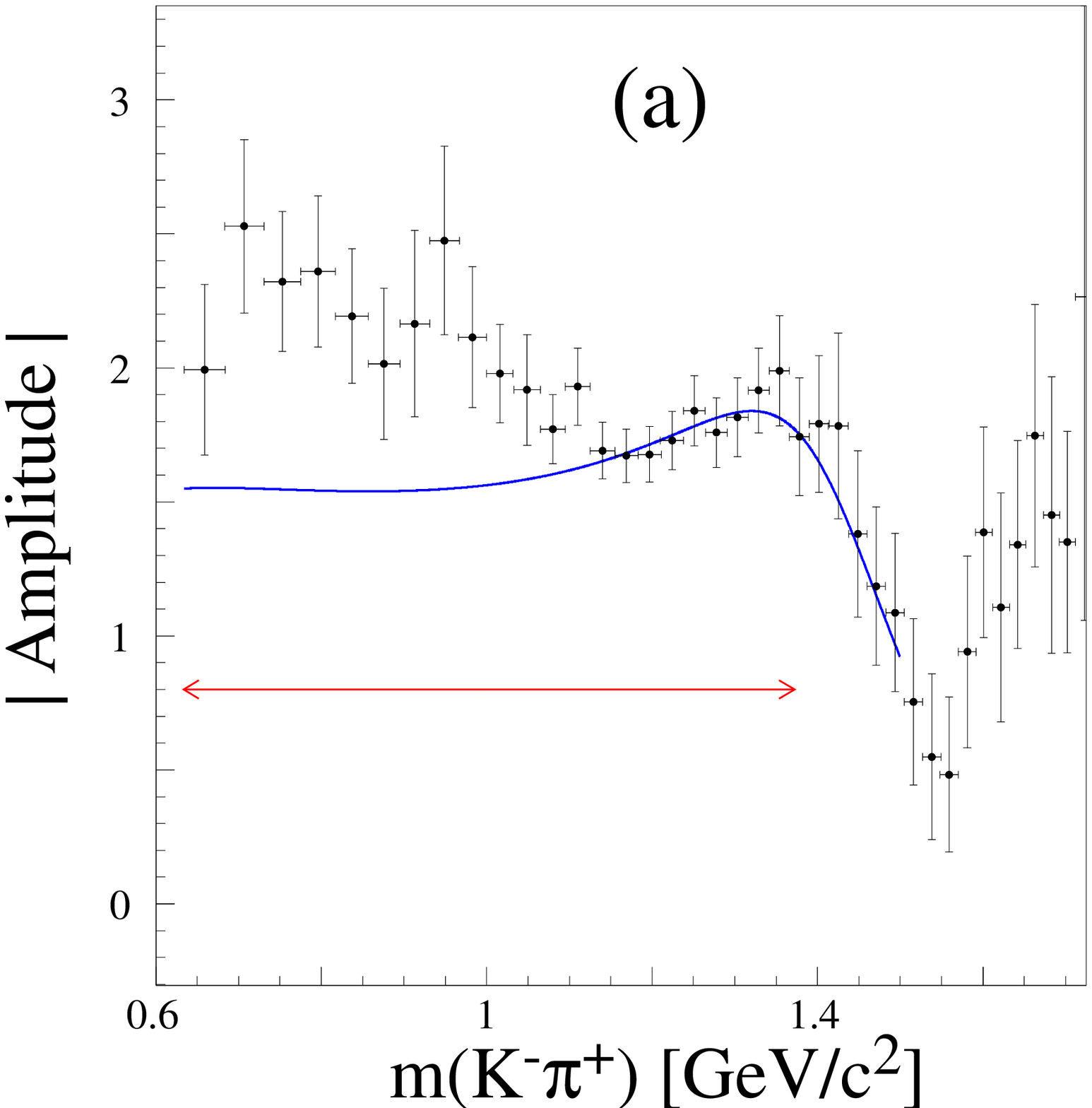}&
\includegraphics[width=0.24\textwidth]{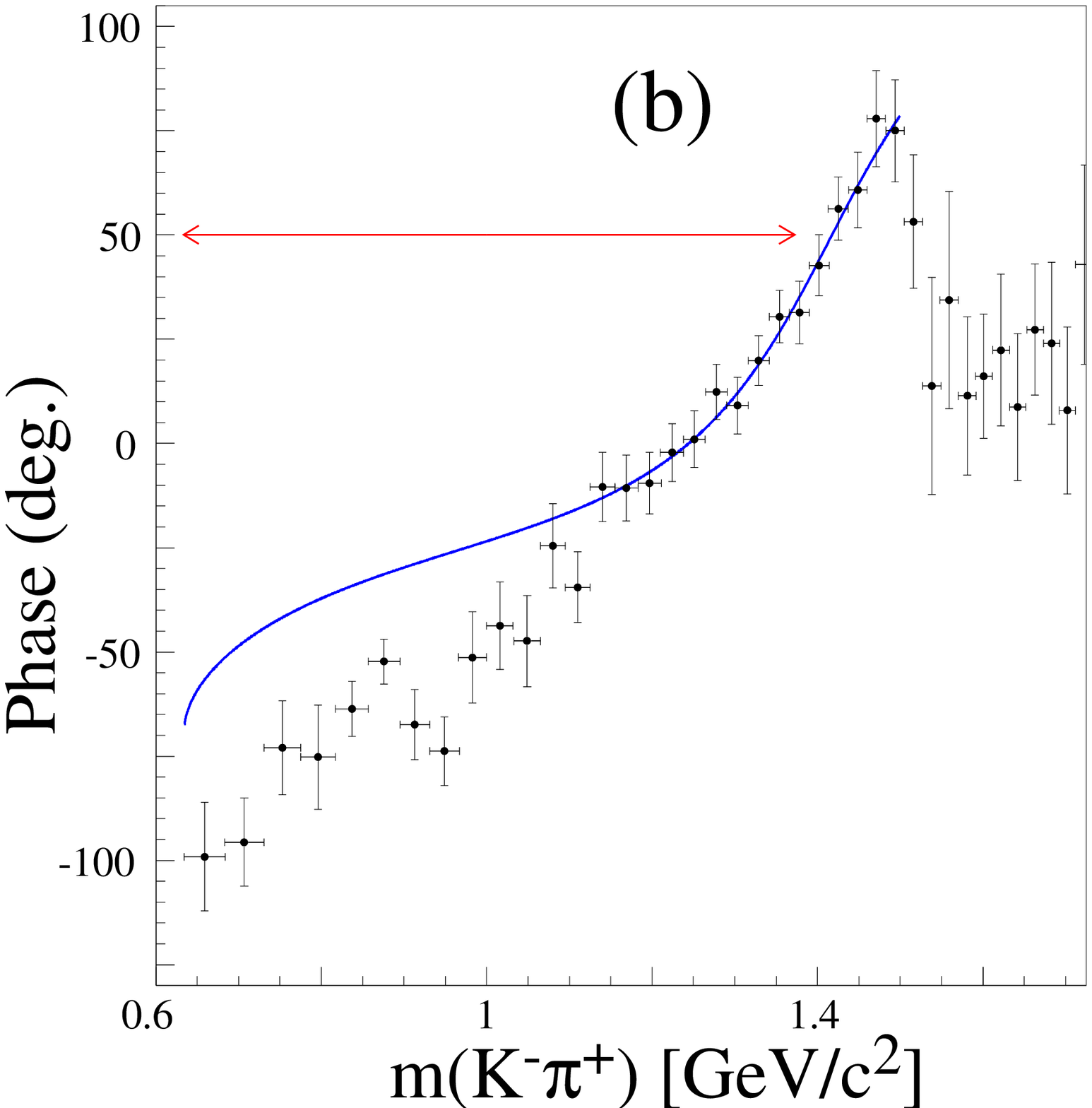} \\
\end{tabular}
\caption{LASS (solid line) and E-791 (dots with error 
bars) $K\pi$ \textit{S-}wave amplitude (a) and phase (b). 
The double headed arrow indicates the mass range available 
in \Dzkkpz.}
\label{Fig2}
\end{figure}

\begin{figure}[!htbp]
\includegraphics[width=3.4in]{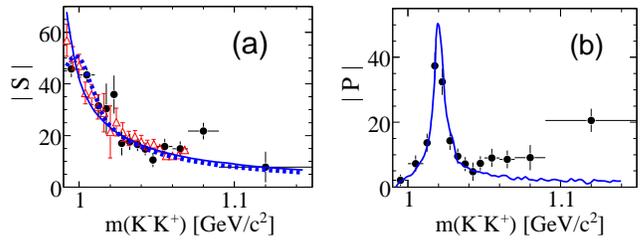}
\vspace{-2em}
\caption{The phase-space-corrected $K^-K^+$ \textit{S-} and 
\textit{P-}wave amplitudes, $\left| S \right|$ and $\left| P \right|$, 
respectively. (a) Lineshapes for (solid line, blue) $f_0(980)$, and (broken 
line, blue) $a_0(980)$. (b) Lineshape for $\phi(1020)$ (solid line, blue). 
In each plot, solid circles with error bars correspond to values obtained from 
the model-independent analysis.
In (a), the open triangles (red) correspond to values obtained from the decay  
$\Dz\to K^-K^+\bar{K^0}$.} 
\label{Fig3}
\end{figure}

\begin{figure*}[!htbp]
\begin{tabular}{cc} 
\includegraphics[width=0.5\textwidth]{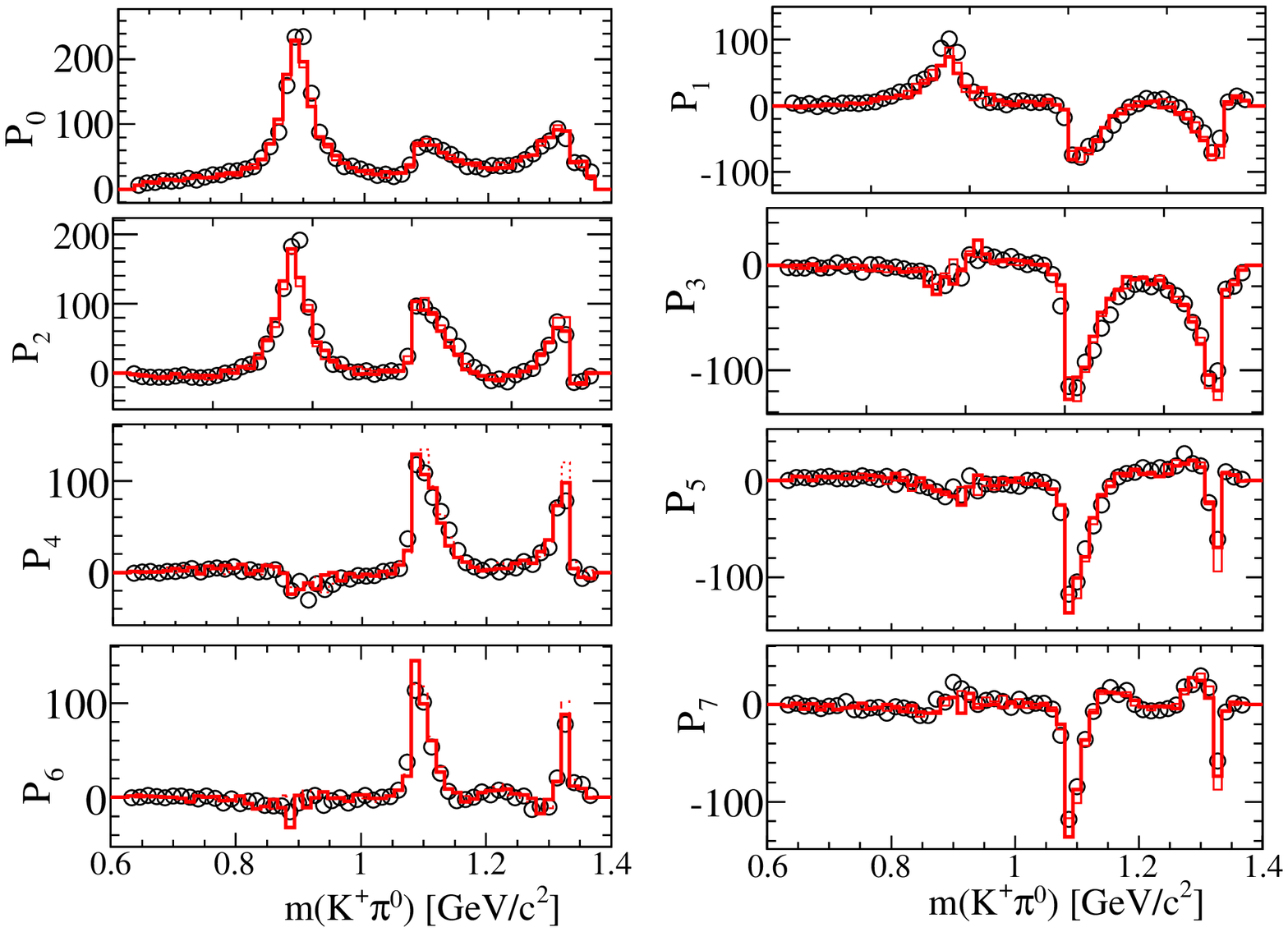} &
\includegraphics[width=0.5\textwidth]{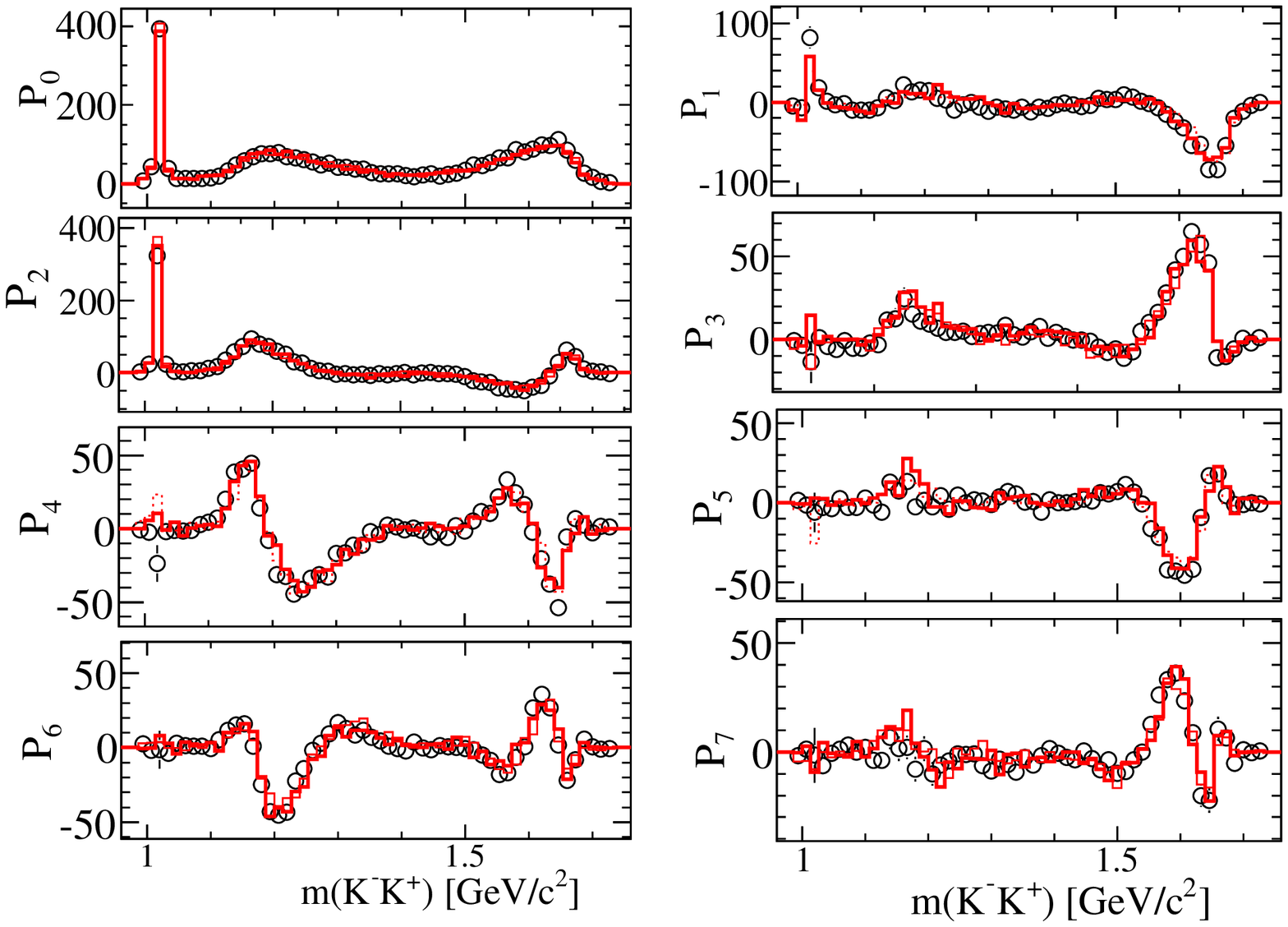} \\
\end{tabular}
\caption{Legendre polynomials moments for the $K^+\piz$ (columns I, II) 
and $K^-K^+$ (columns III, IV) channels of \Dzkkpz. The circles with 
error bars are data points and the curves (red) are derived from the fit 
functions.} 
\label{Fig4}
\end{figure*}

The $K^{\pm}\piz$ systems from the decay 
\Dzkkpz~\cite{note1} can provide information on the $K\pi$ \textit{S-}wave 
amplitude in the mass range 0.6--1.4 \gevcc, and hence on the 
possible existence of the $\kappa(800)$, reported to date only in the neutral 
state ($\kappa^0 \to K^- \pi^+$)~\cite{kappa}. If the $\kappa$ has isospin 
$1/2$, it should be observable also in the charged states. Results of the 
present analysis can be an input for extracting the CKM phase $\gamma$ 
by exploiting interference in the Dalitz 
plot from the decay $B^{\pm}\to\Dz_{K^-K^+\piz}K^{\pm}$~\cite{myGamma}.\\
\indent We perform the analysis on 385 fb${}^{-1}$ data using the same 
event-selection criteria as in our measurement of 
the branching ratio of the decay \Dzkkpz~\cite{mybr}. To minimize uncertainty 
from background shape, we choose a high purity ($\sim$$98\%$) sample  
using $1855 < m_{D^0} < 1875$~\mevcc, and find $11278 \pm 110$ signal 
events. The Dalitz plot for these events is shown in Fig.~\ref{Fig1}(a).

\indent For \Dz\ decays to $K^{\pm}\piz$ \textit{S-}wave states, 
we consider three amplitude models: LASS amplitude for  
$K^-\pi^+\to K^-\pi^+$ elastic scattering~\cite{LASS,mykkpi0}, 
the E-791 results for the $K^-\pi^+$ \textit{S-}wave 
amplitude from a partial-wave analysis of the 
decay $D^+\to K^-\pi^+\pi^+$~\cite{brian}, and a coherent 
sum of a uniform nonresonant term plus Breit-Wigner terms for  
$\kappa(800)$ and $K^*_0(1430)$ resonances.

In Fig.~\ref{Fig2} we compare the $K\pi$ \textit{S-}wave amplitude 
from the E-791 analysis~\cite{brian} to the LASS amplitude.
The LASS $K\pi$ \textit{S-}wave amplitude gives the best agreement 
with data and we use it in our nominal fits ($\chi^2$ probability  62\%). 
The $K\pi$ \textit{S-}wave modeled by the combination of $\kappa(800)$ (with 
parameters taken from Ref.~\cite{kappa}), a nonresonant term and 
$K^*_0(1430)$ has a smaller fit probability ($\chi^2$ probability $<$ 5\%). 
The best fit with this model ($\chi^2$ probability 13\%) yields a charged 
$\kappa$ of mass (870 $\pm$ 30)~\mevcc, and width (150 $\pm$ 20)~\mevcc, 
significantly different from those reported in Ref.~\cite{kappa} for the 
neutral state. This does not support the hypothesis that production of a 
charged, scalar $\kappa$ is being observed. The E-791 amplitude~\cite{brian} 
describes the data well, except near threshold. 
We use it to estimate systematic uncertainty in our results.\\ 
\indent We describe the \Dz decay to a $K^-K^+$ \textit{S-}wave state by a 
coupled-channel Breit-Wigner amplitude for the $f_0(980)$ and $a_0(980)$ 
resonances, with their respective couplings to $\pi\pi$, $K \bar{K}$ and 
$\eta\pi$, $K\bar{K}$ final states~\cite{mykkpi0}.
Only the high mass tails of $f_0(980)$ and $a_0(980)$ are observable, as 
shown in Fig.~\ref{Fig3}.

\begin{table*}[htbp]
\caption{The results obtained from the $D^0 \to K^- K^+ \piz$ Dalitz plot 
fit~\cite{mykkpi0}. The errors are statistical and systematic, 
respectively. We show the $a_0(980)$ contribution, when it is included in 
place of the $f_0(980)$, in square brackets.}
\label{tab:result}
\begin{tabular}{|l|rrr|rrr|}
\hline
          &&   Model I &   &&  Model II & \cr
\hline
State &   Amplitude, $a_r$ &  Phase, $\phi_r$ (${}^\circ$) & Fraction, 
 $f_r$ (\%) &   Amplitude, $a_r$ &  Phase, $\phi_r$ (${}^\circ$) &  
 Fraction, $f_r$ (\%)\cr 
 \hline\hline
$K^*(892)^{+}$ & 1.0 (fixed) & 0.0 (fixed) & 45.2$\pm$0.8$\pm$0.6 & 1.0 (fixed)
& 0.0 (fixed) &  44.4$\pm$0.8$\pm$0.6\cr
$K^*(1410)^{+}$ &  2.29$\pm$0.37$\pm$0.20  &   86.7$\pm$12.0$\pm$9.6  
&  3.7$\pm$1.1$\pm$1.1 &   &    & \cr
$K^+\piz(\textit{S})$  &  1.76$\pm$0.36$\pm$0.18 & -179.8$\pm$21.3$\pm$12.3 
&  16.3$\pm$3.4$\pm$2.1&  3.66$\pm$0.11$\pm$0.09  & -148.0$\pm$2.0$\pm$2.8   
&  71.1$\pm$3.7$\pm$1.9\cr
$\phi(1020)$ &  0.69$\pm$0.01$\pm$0.02  &  -20.7$\pm$13.6$\pm$9.3  
&  19.3$\pm$0.6$\pm$0.4 &  0.70$\pm$0.01$\pm$0.02  
&   18.0$\pm$3.7$\pm$3.6   &  19.4$\pm$0.6$\pm$0.5\cr
$f_0(980)$ &  0.51$\pm$0.07$\pm$0.04  & -177.5$\pm$13.7$\pm$8.6  
&  6.7$\pm$1.4$\pm$1.2 &  0.64$\pm$0.04$\pm$0.03  &  -60.8$\pm$2.5$\pm$3.0 
&  10.5$\pm$1.1$\pm$1.2\cr
$\left[a_0(980)^0\right]$ & [0.48$\pm$0.08$\pm$0.04]& [-154.0$\pm$14.1$\pm$8.6]
&  [6.0$\pm$1.8$\pm$1.2] &  [0.68$\pm$0.06$\pm$0.03]& [-38.5$\pm$4.3$\pm$3.0] 
&  [11.0$\pm$1.5$\pm$1.2]\cr
$f_2'(1525)$  &  1.11$\pm$0.38$\pm$0.28  &  -18.7$\pm$19.3$\pm$13.6 &  
0.08$\pm$0.04$\pm$0.05 &   &  &  \cr
$K^*(892)^{-}$ &  0.601$\pm$0.011$\pm$0.011 &  -37.0$\pm$1.9$\pm$2.2   
&  16.0$\pm$0.8$\pm$0.6 &  0.597$\pm$0.013$\pm$0.009 
&  -34.1$\pm$1.9$\pm$2.2   &  15.9$\pm$0.7$\pm$0.6\cr
$K^*(1410)^{-}$ &  2.63$\pm$0.51$\pm$0.47  & -172.0$\pm$6.6$\pm$6.2   
&  4.8$\pm$1.8$\pm$1.2 &   &    & \cr
$K^-\piz(\textit{S})$& 0.70$\pm$0.27$\pm$0.24 & 133.2$\pm$22.5$\pm$25.2 
& 2.7$\pm$1.4$\pm$0.8 &  0.85$\pm$0.09$\pm$0.11  &  108.4$\pm$7.8$\pm$8.9 
& 3.9$\pm$0.9$\pm$1.0\cr
\hline 
\end{tabular}
\end{table*}

\begin{figure}[!htbp]
\includegraphics[width=0.5\textwidth]{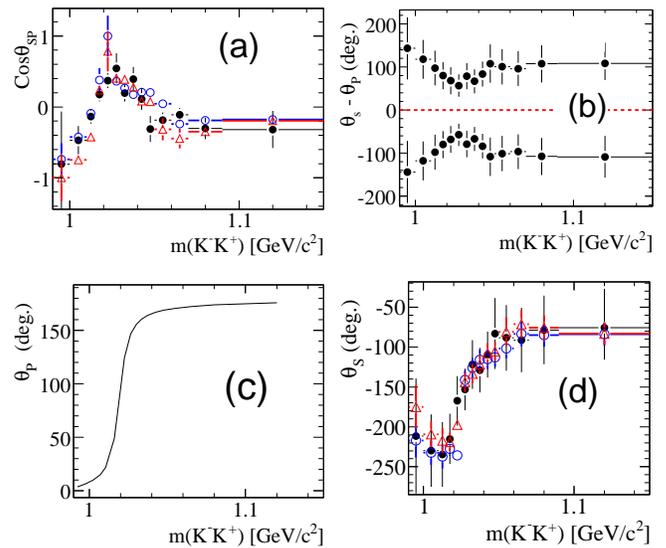}
\caption{Results of the partial-wave analysis of the $K^-K^+$ 
system. (a) Cosine of relative
phase $\theta_{SP}=\theta_\textit{S} - \theta_P$, (b) two solutions for 
$\theta_{SP}$, (c) \textit{P-}wave phase for $\phi(1020)$, and (d) 
\textit{S-}wave phase derived from the upper solution in (b). Solid bullets 
are data points, and open circles (blue) and open triangles (red) correspond, 
respectively, to isobar models I, II.}
\label{Fig5}
\end{figure}

\indent We find that two different isobar models describe the data well. Both 
yield almost identical behavior in invariant mass 
(Fig.~\ref{Fig1}b--\ref{Fig1}d) and angular distribution (Fig.~\ref{Fig4}). 
The dominance of $\Dz\to K^{*+}K^-$ over $\Dz\to K^{*-}K^+$ suggests that, in 
tree-level diagrams, the form factor for \Dz\ coupling to $K^{*-}$ is 
suppressed compared to the corresponding $K^-$ coupling. While the measured 
fit fraction for $\Dz\to K^{*+}K^-$ agrees well with a phenomenological 
prediction~\cite{theory} based on a large SU(3) symmetry breaking, the 
corresponding results for $\Dz\to K^{*-}K^+$ and the color-suppressed 
$\Dz\to\phi\pi^0$ decays differ significantly.
It appears from Table~\ref{tab:result}
that the $K^+\pi^0$ \textit{S-}wave amplitude can absorb any $K^*(1410)$ and   
$f_2'(1525)$ if those are not in the model. The other components are quite 
well established, independent of the model.
From Table~\ref{tab:result}, the strong phase difference, $\delta_D$, 
between the \Dzb and \Dz decays to $K^*(892)^{+}K^-$ state and their amplitude 
ratio, $r_D$, are given by: $\delta_D$ = 
$-35.5^\circ \pm 1.9^\circ$ (stat) $\pm 2.2^\circ$ (syst) and $r_D$ = 0.599 
$\pm$ 0.013 (stat) $\pm$ 0.011 (syst)~\cite{mykkpi0}.
Systematic uncertainties in quantities in Table~\ref{tab:result} 
arise from experimental effects (\textit{e.g.,} efficiency parameters, 
background shape, particle-identification), and also 
from uncertainty in the nature of the models used to describe the 
data (\textit{e.g.,} $K\pi$ \textit{S-}wave amplitude and resonance 
parameters).\\
\indent We show the Legendre polynomials moments in 
Fig.~\ref{Fig4} for the $K^+ \piz$ and $K^- K^+$ channels, for $l = 0-7$.
We use the relations of Eq.~\ref{eq:pwa} to evaluate 
$\left| S \right|$  and $\left| P \right|$ shown in Fig.~\ref{Fig3}, 
and $\theta_\textit{SP}$ shown in Fig.~\ref{Fig5}, for the 
$K^-K^+$ channel in the mass range $m_{K^-K^+} < 1.15~\gevcc$. The measured 
values of $\left| S \right|$ agree well with those obtained in the analysis 
of the decay $\Dz\to K^-K^+\bar{K^0}$~\cite{antimo} and also with 
either the $f_0(980)$ or the $a_0(980)$ lineshape. The measured values of 
$\left| P \right|$ are consistent with a Breit-Wigner lineshape for 
$\phi(1020)$.

\section{Dalitz plot analysis of \Dzpppz}

\begin{figure}[!htbp]
\begin{tabular}{cc} 
\includegraphics[width=0.235\textwidth]{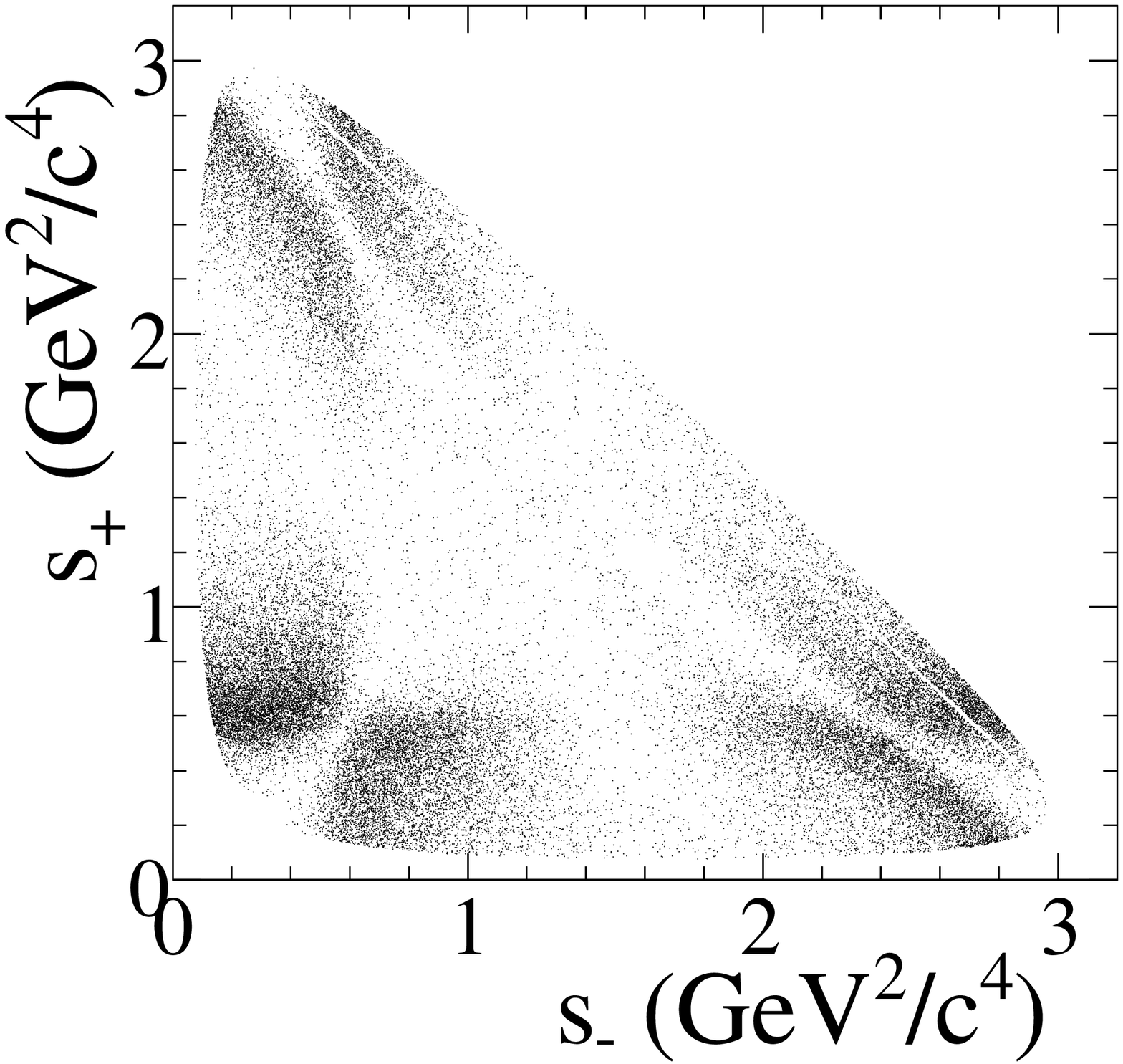}&
\includegraphics[width=0.235\textwidth]{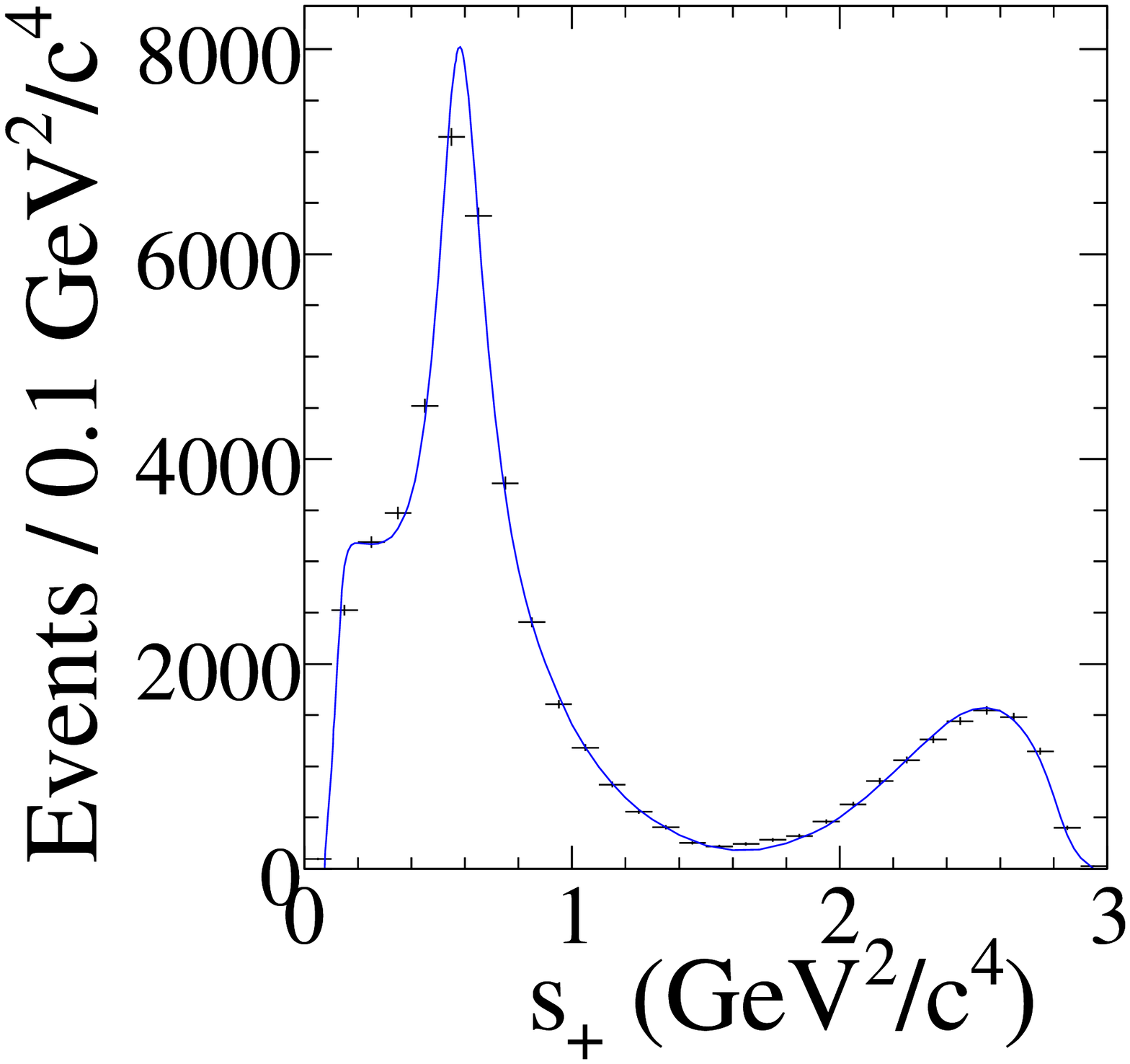}\\
\end{tabular}
\begin{tabular}{cc} 
\includegraphics[width=0.235\textwidth]{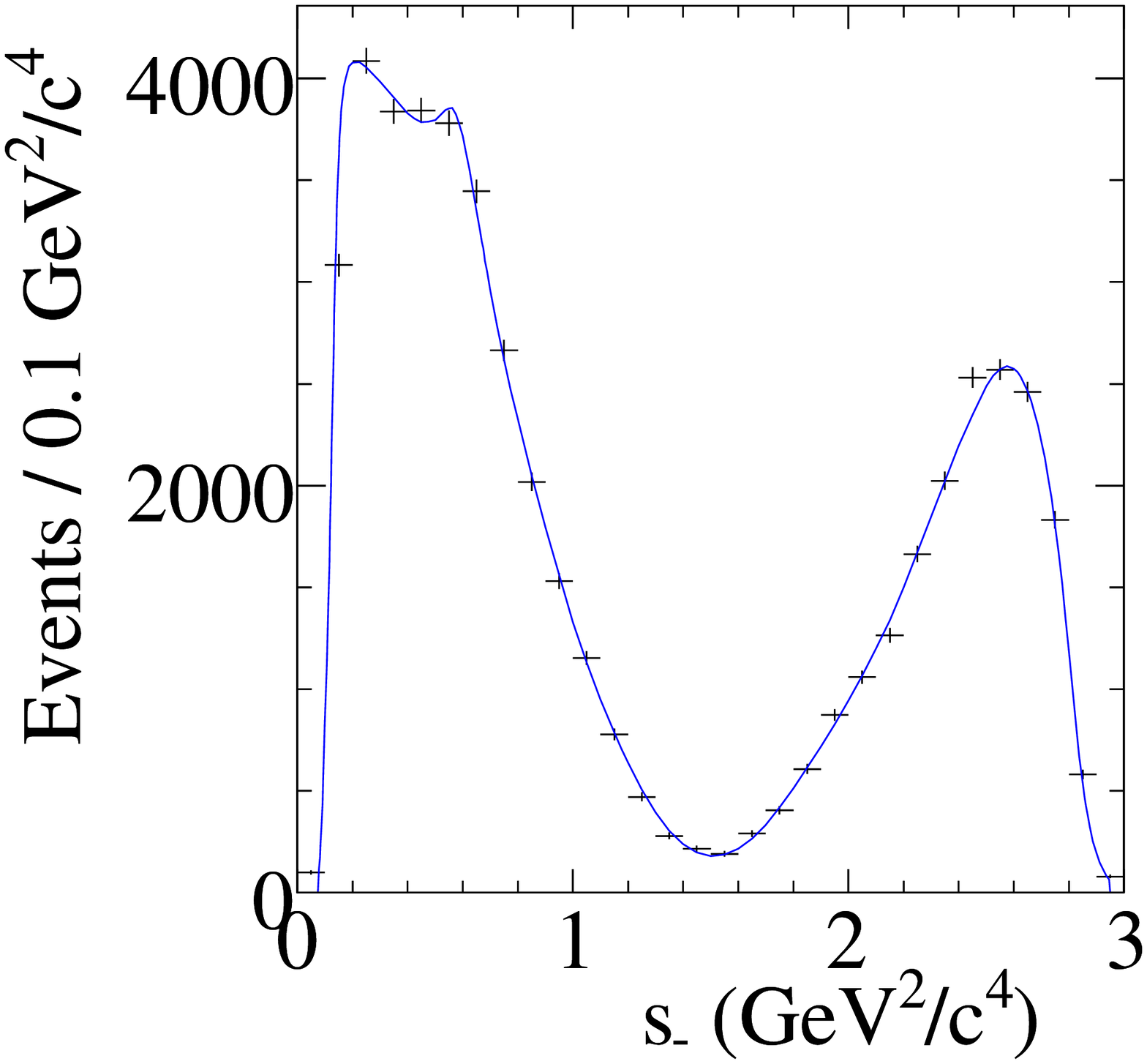}&
\includegraphics[width=0.235\textwidth]{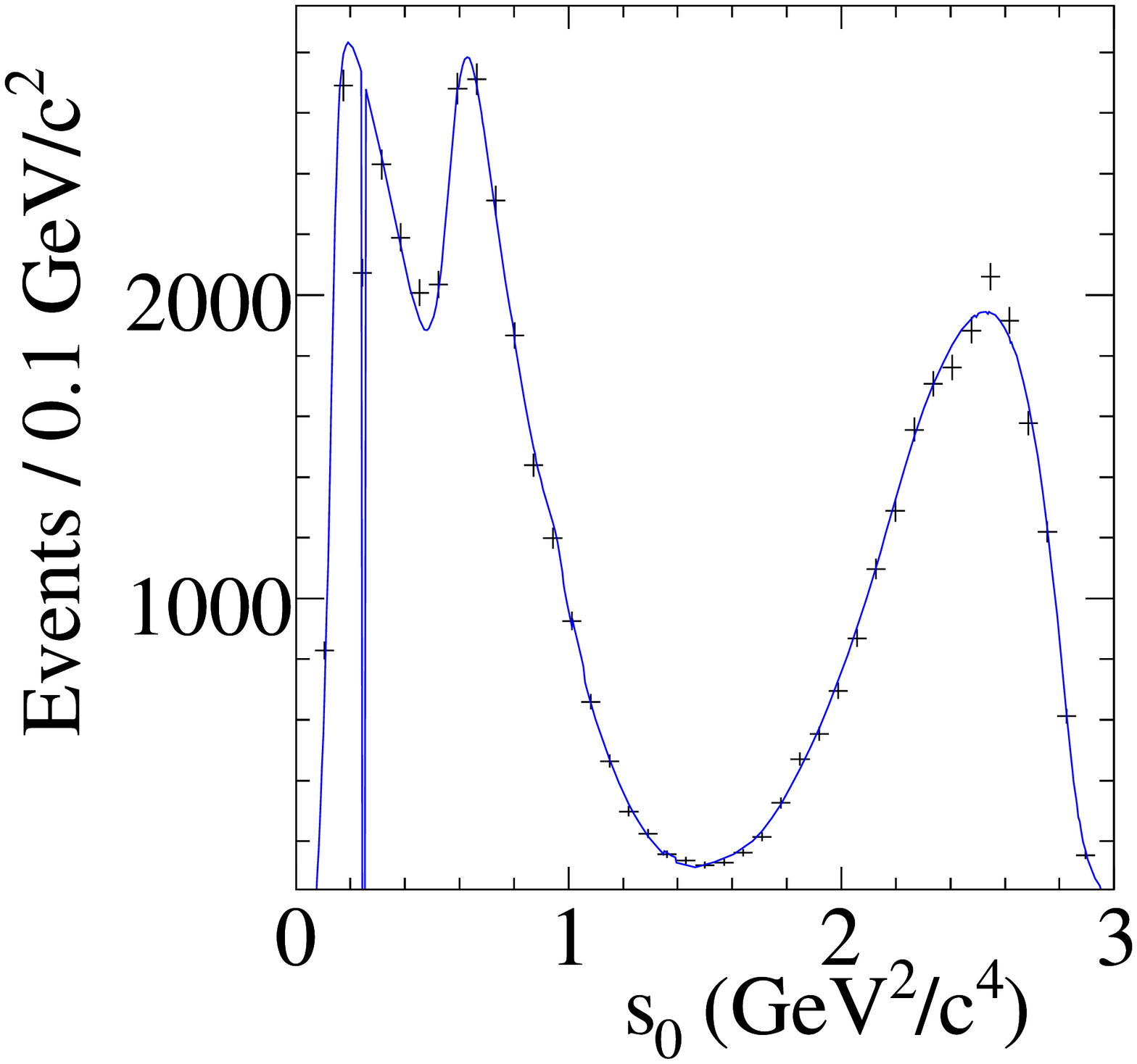}\\
\end{tabular}
\caption{Dalitz plot and invariant mass-squared 
projections for the $\Dzpppz$ decay excluding 
$\Dz\to K_s^0\piz$.} 
\label{Fig6}
\end{figure}

An important component 
of the program to study \CP~violation is the measurement of
the angle $\gamma$ of the unitarity triangle related to the 
Cabibbo-Kobayashi-Maskawa quark mixing matrix.
The decays $\btodkgen$ 
can be used to measure $\gamma$ with
essentially no hadronic uncertainties, exploiting interference
between \btou\ and \btoc\ decay amplitudes. 
The most effective method to measure $\gamma$ has turned out to be 
the analysis of the $D$-decay Dalitz plot
distribution in  $\bpmtodkpm$ with multi-body $D$ decays~\cite{ref:D}.
This method has only been used with the Cabibbo-favored decay $D\to
\KS \pi^+\pi^-$~\cite{Abe:2003cn,Aubert:2004kv}.
We perform the first \CP-violation study of
$\bpmtodkpm$ using a multibody, Cabibbo-suppressed $D$ decay,	
$\dtoppp$. 

\begin{figure*}[!htbp]
\begin{tabular}{cc} 
\includegraphics[width=0.5\textwidth]{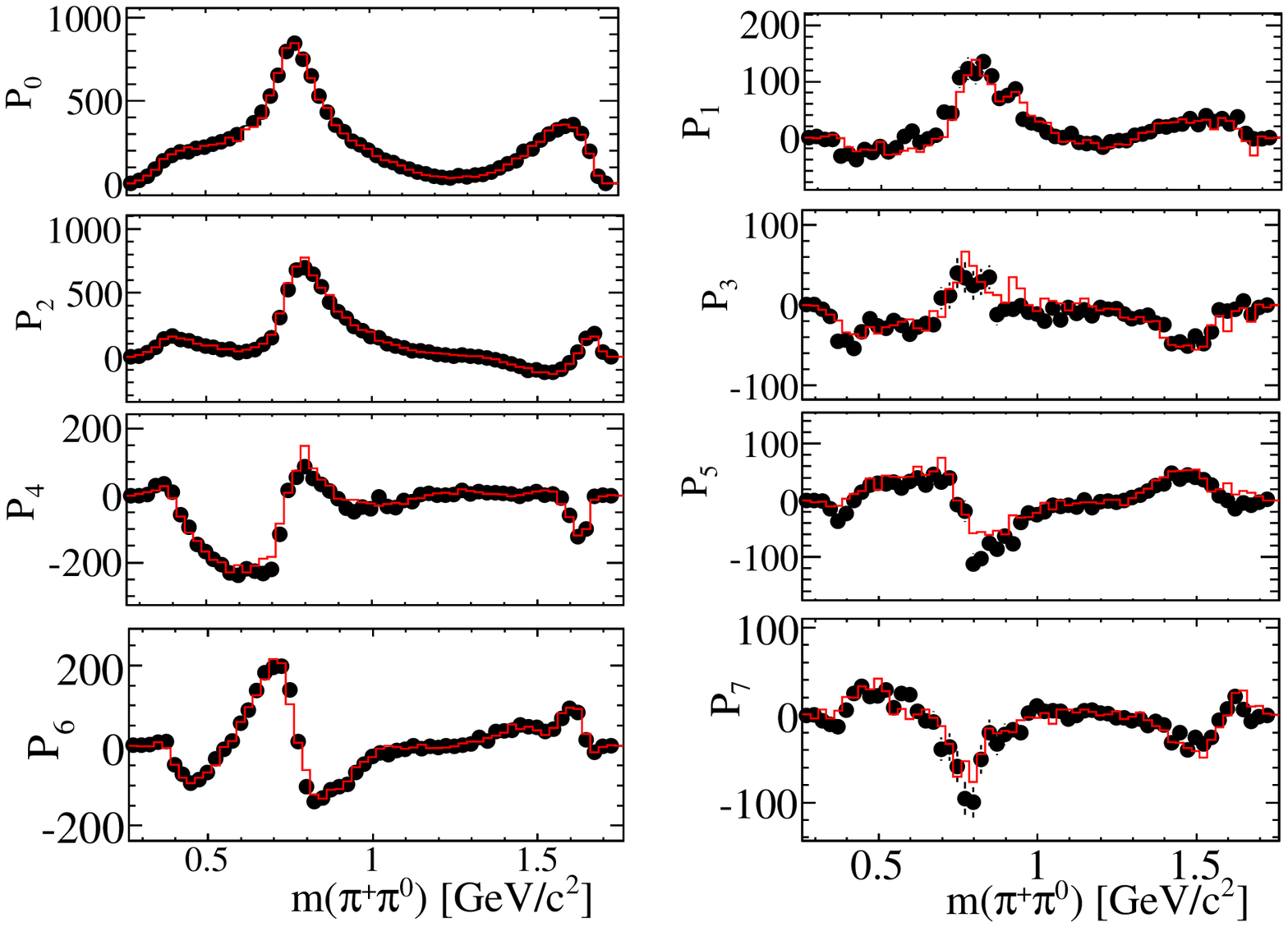} &
\includegraphics[width=0.5\textwidth]{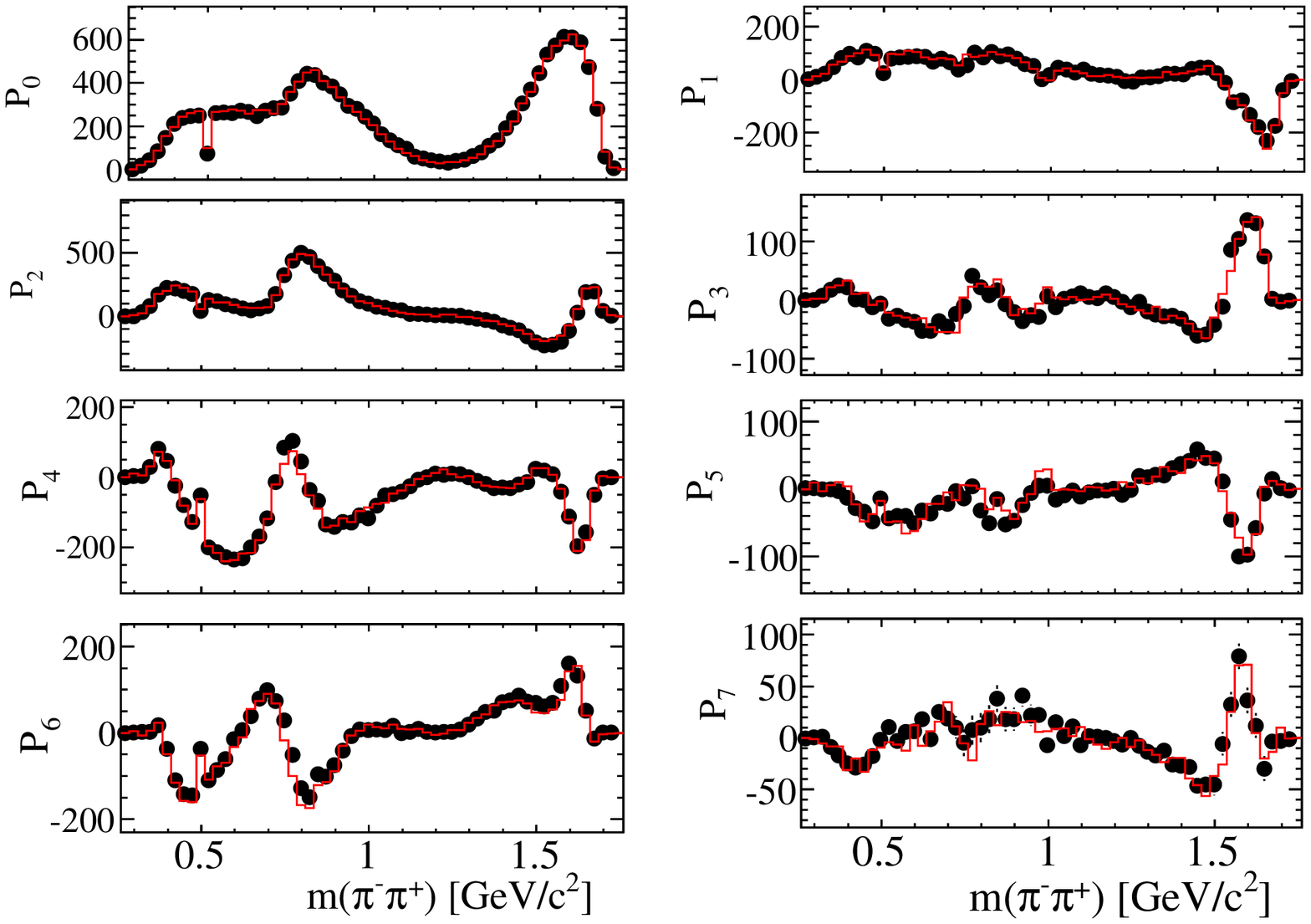}\\ 
\end{tabular}
\caption{Legendre polynomials moments for the $\pi^+\piz$ (columns I, II) 
and $\pi^-\pi^+$ (columns III, IV) channels of \Dzpppz. The circles with 
error bars are data points and the curves (red) are derived from the fit 
functions.} 
\label{Fig7}
\end{figure*}

We determine the parameters $a_r$, $\phi_r$, and $f_r$
by fitting a large sample of $\Dz$ and $\Dzb$ mesons,
flavor-tagged through their production in the decay
$D^{*+}\to\Dz\pi^+$~\cite{mybr}. 
Of the $D$ candidates in the signal region $1848 < m_{\Dz} < 1880$~\mevcc,
we obtain from the fit $44780 \pm 250$ signal 
and $830 \pm 70$ background events.

Table~\ref{tab:DstarFit} summarizes the results of this fit, with
systematic errors
obtained by varying the masses and
widths of the $\rho(1700)$ and $\sigma$ resonances and the form factors, 
and also varying the signal efficiency parameters 
to account for uncertainties in reconstruction and particle identification.
The Dalitz plot
distribution of the data is shown in
Fig.~\ref{Fig6}(a-d). 
The distribution is marked by 
three destructively interfering $\rho\pi$ amplitudes, 
suggesting a final state dominated by $I=0$~\cite{ref:zemach}.
We show the Legendre polynomials moments in 
Fig.~\ref{Fig7} for the $\pi^+ \piz$ and $\pi^- \pi^+$ channels, for $l = 0-7$.
The agreement between data and fit is again excellent.
Unlike in case of the decay \Dzkkpz, we cannot 
use the relations of Eq.~\ref{eq:pwa} to evaluate 
$\left| S \right|$  and $\left| P \right|$,  and 
$\theta_\textit{SP}$ in any of the two-body $\pi\pi$ channels because of 
the contributions from cross-channels in the entire available mass-range.  

\begin{figure}[!htbp]
\begin{tabular}{cc} 
\includegraphics[width=0.235\textwidth]{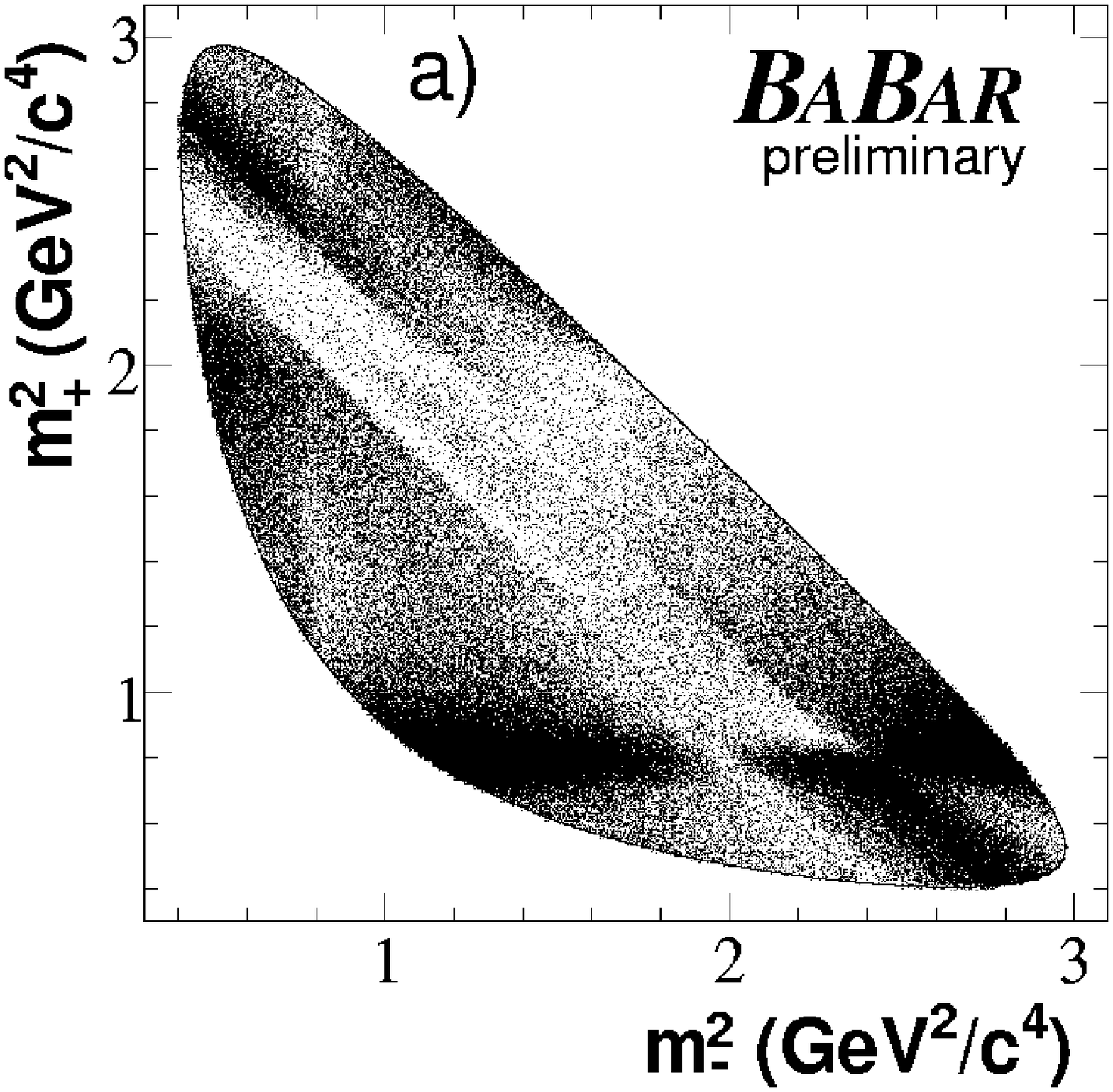}&
\includegraphics[width=0.235\textwidth]{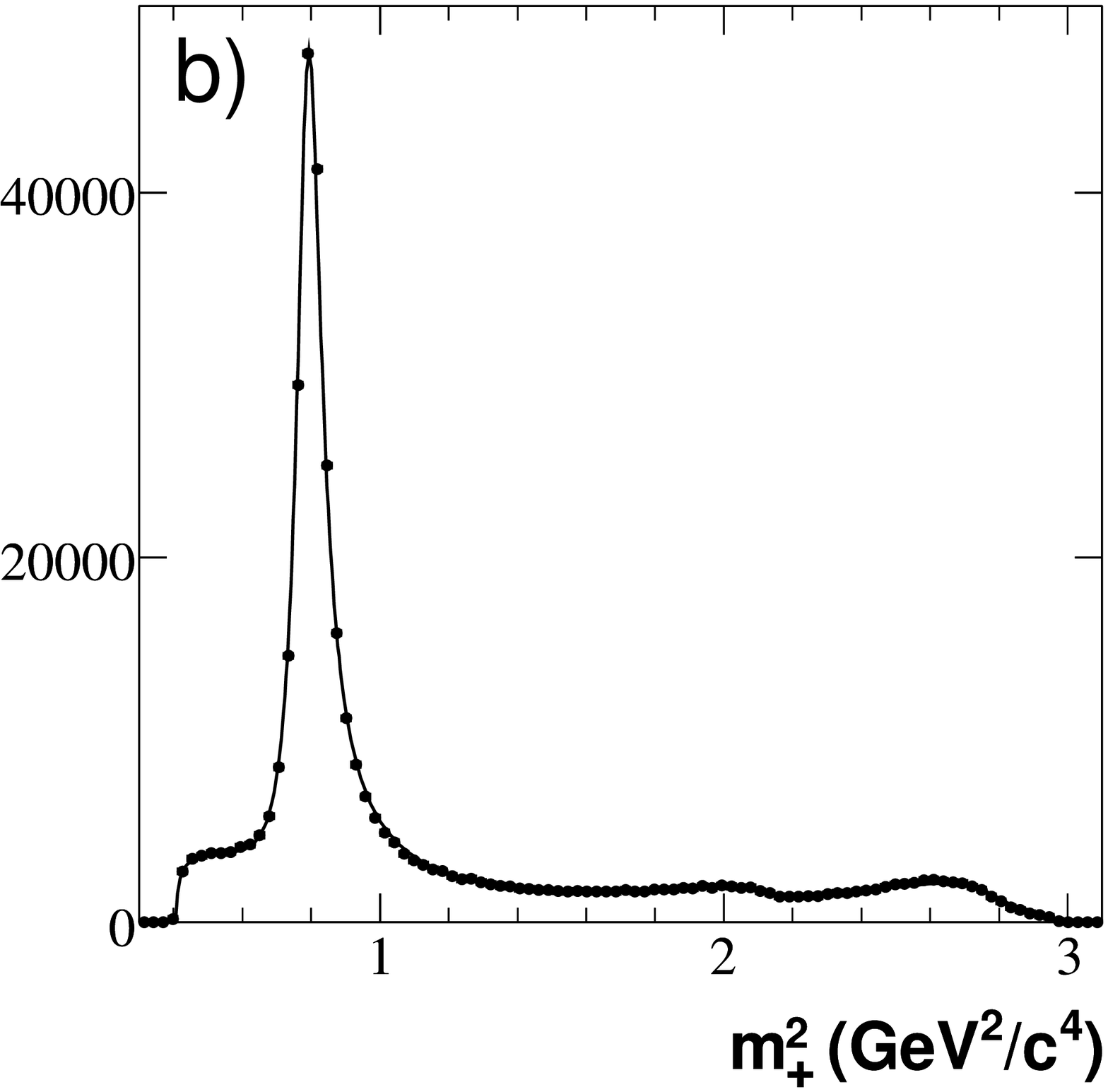}\\
\end{tabular}
\begin{tabular}{cc} 
\includegraphics[width=0.235\textwidth]{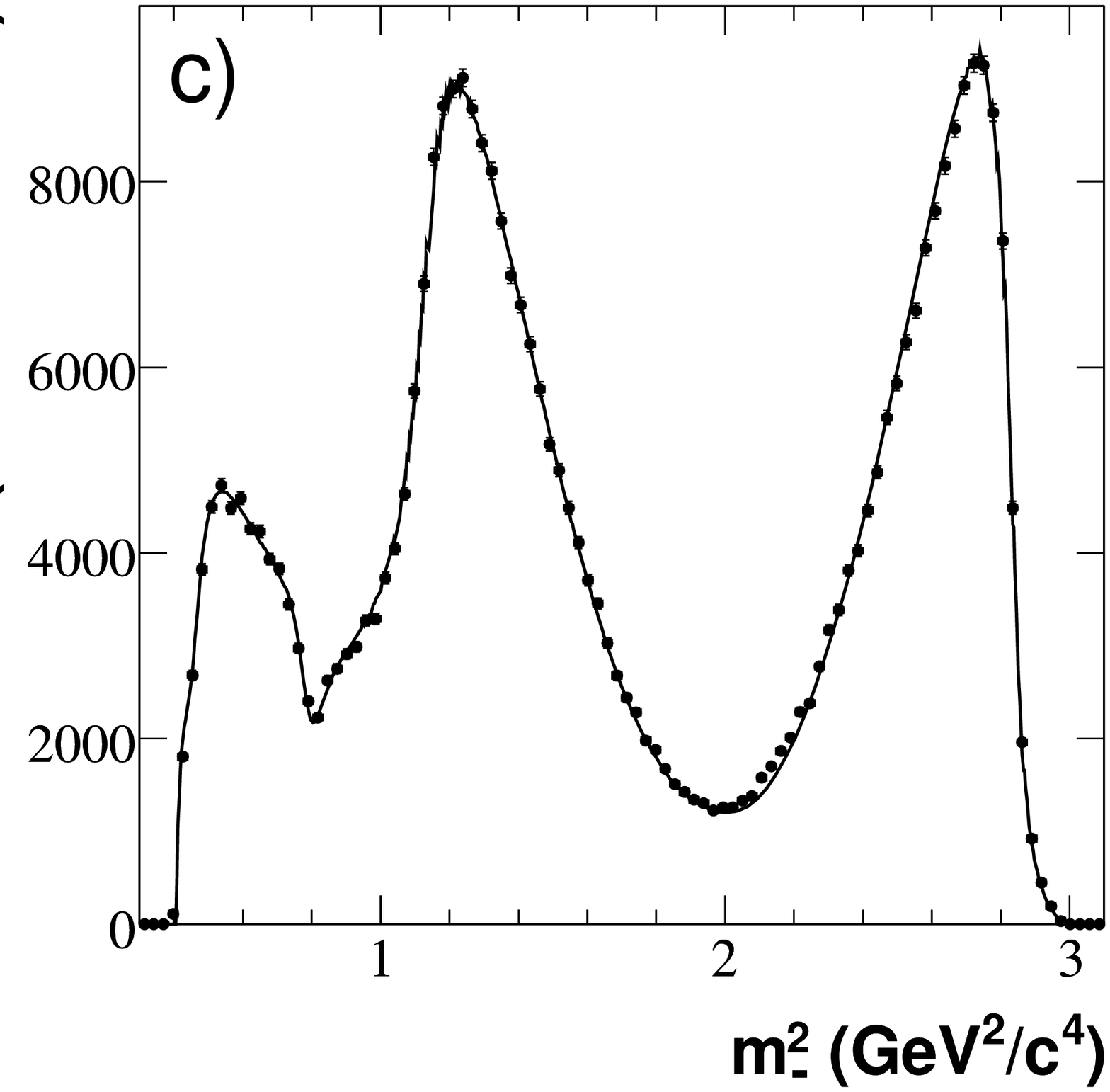}&
\includegraphics[width=0.235\textwidth]{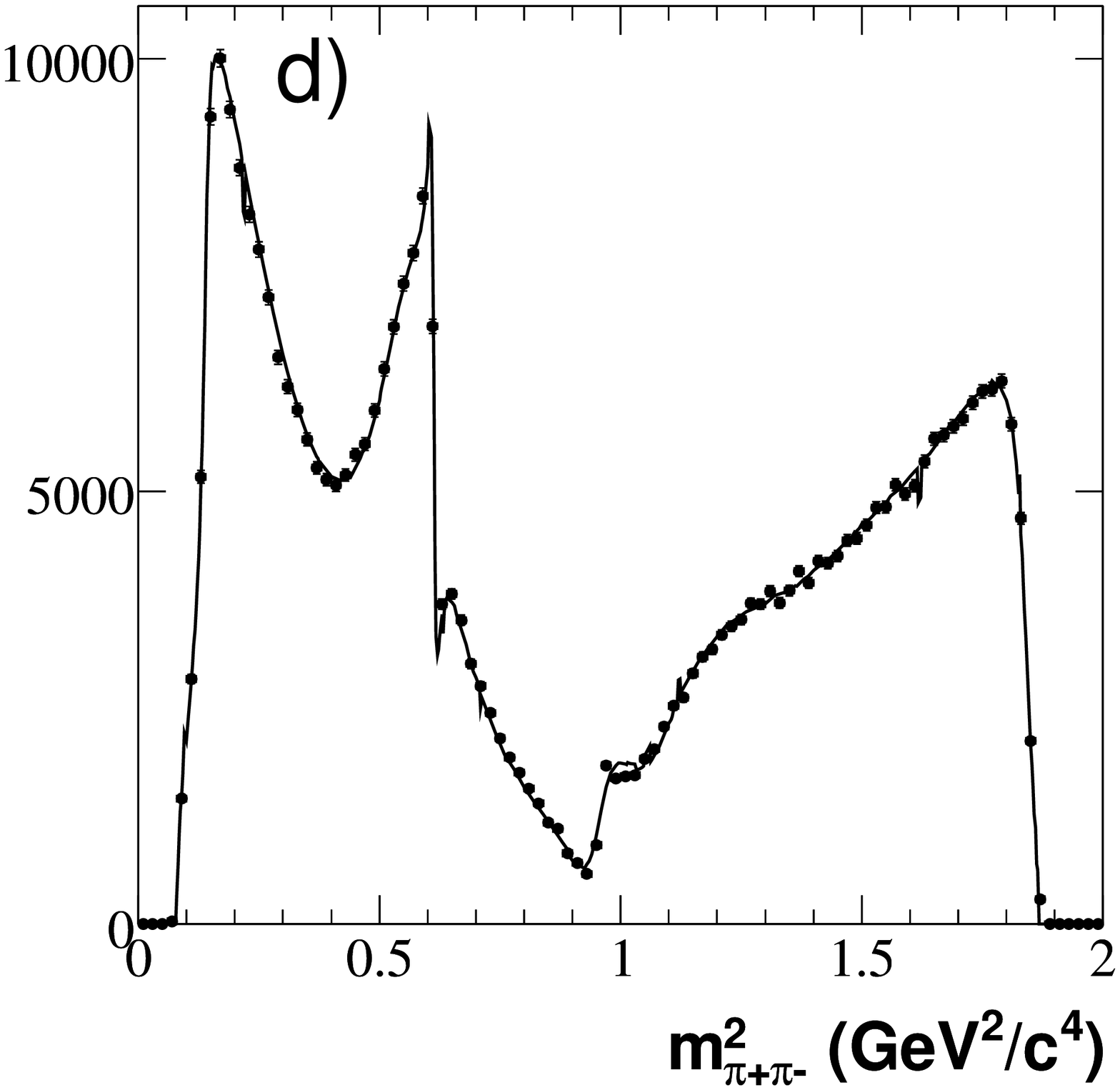}\\
\end{tabular}
\caption{(a) The $\bar{D}^0 \to \KS \pi^- \pi^+$ Dalitz distribution 
from $D^{*-} \to \bar{D}^0 \pi^-$ events, and projections
on (b) $m^2_+=m^2_{\KS\pi^+}$, (c) $m^2_-=m^2_{\KS\pi^-}$, and 
(d) $m^2_{\pi^+\pi^-}$. $\Dz \to \KS \pi^+ \pi^-$ from 
$D^{*+} \to \Dz \pi^+$ events are also included. The curves are the 
model fit projections.}
\label{Fig8}
\end{figure}

\begin{figure}[!htbp]
\includegraphics[width=0.48\textwidth]{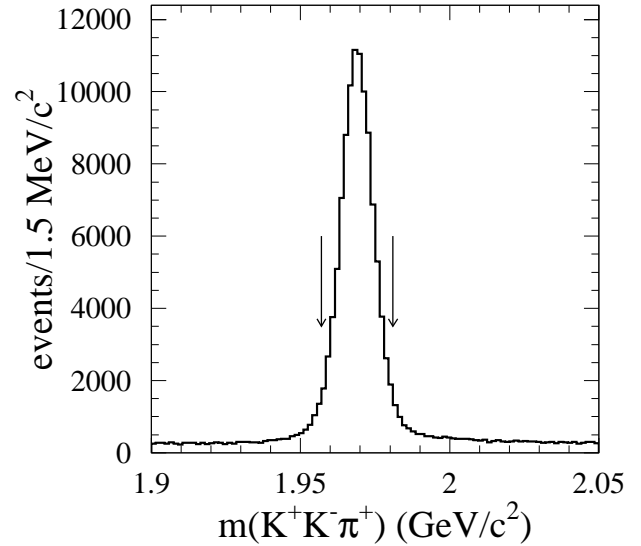}
\caption{The invariant mass distribution of the reconstructed $D_S$ 
candidate in the decay $\Dskkp$. For the Dalitz plot analysis we use 
events in the mass window shown by vertical arrows. The results are 
preliminary.}
\label{Fig9}
\end{figure}

\begin{figure}[!htbp]
\includegraphics[width=0.48\textwidth]{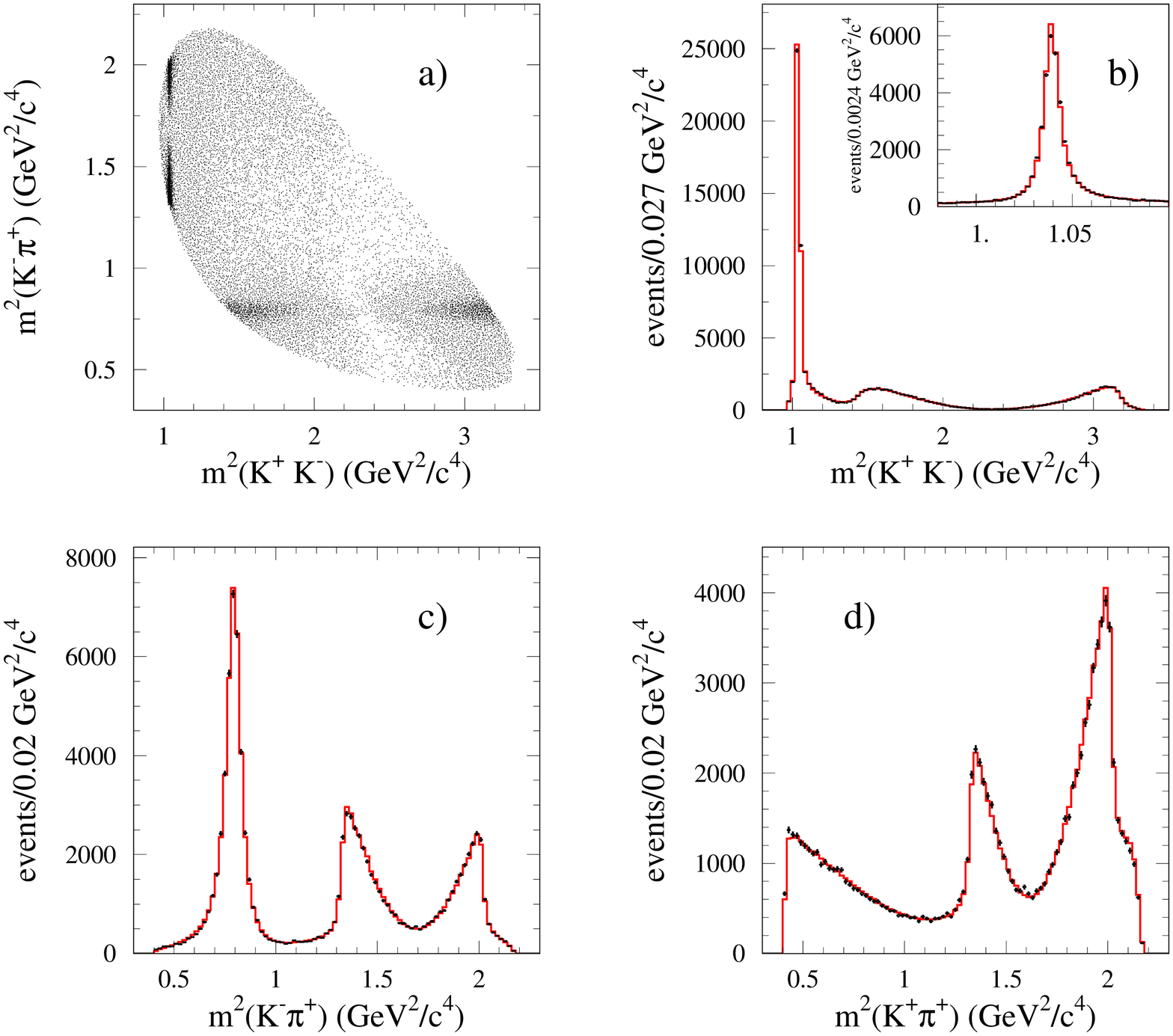}
\caption{(a) The $\Dskkp$ Dalitz distribution, and projections
on (b) $m^2_{K^+K^-}$, (c) $m^2_{K^-\pi^+}$, and 
(d) $m^2_{K^+\pi^+}$. The curves are the 
model fit projections. The results are preliminary.}
\label{Fig10}
\end{figure}

\begin{figure*}[tb]
\includegraphics[width=0.98\textwidth]{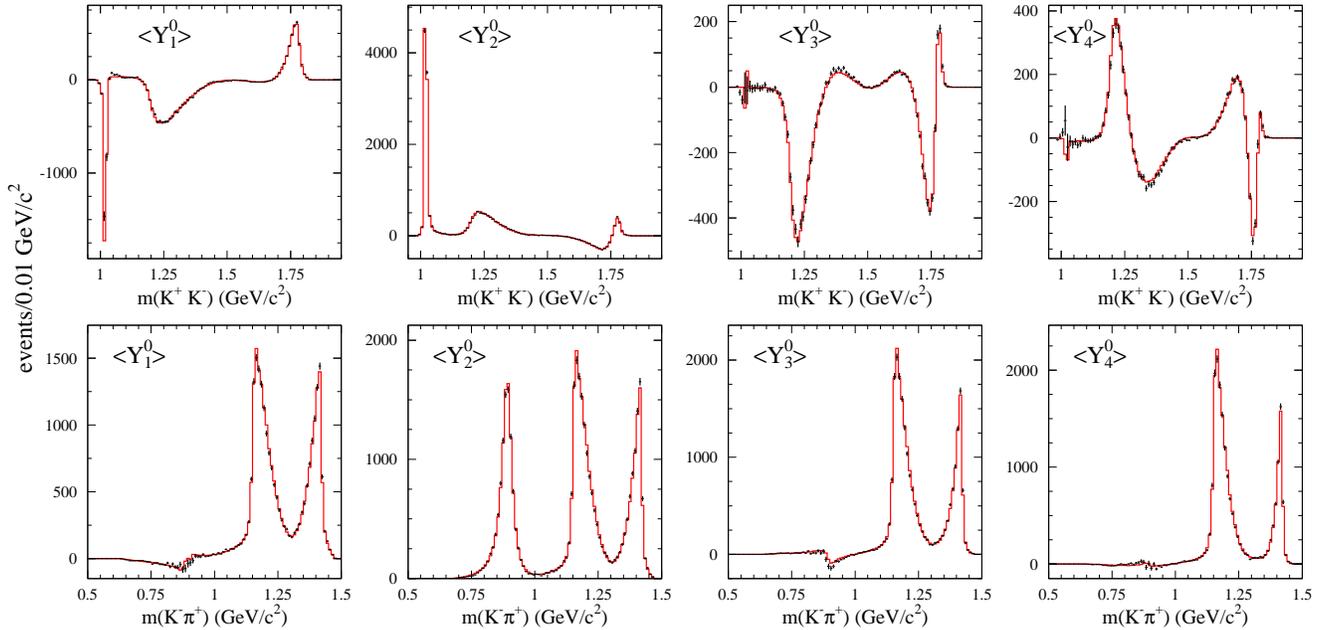}
\caption{Legendre polynomials moments for the $K^+K^-$ (top) 
and $K^-\pi^+$ (bottom) channels of \Dskkp. The dots with 
error bars are data points and the curves are derived from the fit 
functions. The results are preliminary.}
\label{Fig11}
\end{figure*}

\begin{table}[!htbp]
\caption{\label{tab:DstarFit}
The results obtained from the \Dzpppz\ Dalitz plot 
fit~\cite{myGamma}. The errors are statistical and systematic, 
respectively. We take the mass (width) of the 
$\sigma$ meson to be 400 (600)~\mevcc.}
   \centering
   \begin{tabular}{l|r|r|r}
     \hline\hline
     State & $a_r$ (\%) & $\phi_r$ ($^\circ$) & $f_r (\%)$\cr
     \hline
   $\rho(770)^+$  &   100 &  0 &  67.8$\pm$0.0$\pm$0.6\cr
     $\rho(770)^0$  &   58.8$\pm$0.6$\pm$0.2 &  16.2$\pm$0.6$\pm$0.4 & 26.2$\pm$0.5$\pm$1.1\cr
     $\rho(770)^-$  &   71.4$\pm$0.8$\pm$0.3 & $-$2.0$\pm$0.6$\pm$0.6  & 34.6$\pm$0.8$\pm$0.3\cr
     $\rho(1450)^+$ &   21$\pm$6$\pm$13    &  $-$146$\pm$18$\pm$24    & 0.11$\pm$0.07$\pm$0.12\cr
     $\rho(1450)^0$ &   33$\pm$6$\pm$4    &  10$\pm$8$\pm$13       & 0.30$\pm$0.11$\pm$0.07\cr
     $\rho(1450)^-$ &   82$\pm$5$\pm$4    &  16$\pm$3$\pm$3       & 1.79$\pm$0.22$\pm$0.12\cr
     $\rho(1700)^+$ &   225$\pm$18$\pm$14    & $-$17$\pm$2$\pm$3       & 4.1$\pm$0.7$\pm$0.7\cr
     $\rho(1700)^0$ &   251$\pm$15$\pm$13    & $-$17$\pm$2$\pm$2       & 5.0$\pm$0.6$\pm$1.0\cr
     $\rho(1700)^-$ &   200$\pm$11$\pm$7    & $-$50$\pm$3$\pm$3       & 3.2$\pm$0.4$\pm$0.6\cr
     $f_0(980)$ & 1.50$\pm$0.12$\pm$0.17 & $-$59$\pm$5$\pm$4 & 0.25$\pm$0.04$\pm$0.04\cr
     $f_0(1370)$ & 6.3$\pm$0.9$\pm$0.9 & 156$\pm$9$\pm$6 & 0.37$\pm$0.11$\pm$0.09\cr
     $f_0(1500)$ & 5.8$\pm$0.6$\pm$0.6 & 12$\pm$9$\pm$4 & 0.39$\pm$0.08$\pm$0.07\cr
     $f_0(1710)$ &11.2$\pm$1.4$\pm$1.7 & 51$\pm$8$\pm$7 & 0.31$\pm$0.07$\pm$0.08\cr
     $f_2(1270)$ & 104$\pm$3$\pm$21 & $-$171$\pm$3$\pm$4 & 1.32$\pm$0.08$\pm$0.10\cr
     $\sigma(400)$ & 6.9$\pm$0.6$\pm$1.2 & 8$\pm$4$\pm$8 & 0.82$\pm$0.10$\pm$0.10\cr
    Non-Res       &   57$\pm$7$\pm$8    &  $-$11$\pm$4$\pm$2      & 0.84$\pm$0.21$\pm$0.12\cr
     \hline\hline
   \end{tabular}
\end{table}

\section{Dalitz plot analysis of \Dzkspp}

The Dalitz plot analysis of the decay \Dzkspp\ is also motivated by its application 
to the measurement of CKM phase $\gamma$~\cite{babar_summer06}. 
We determine the $\Dz \to \KS \pi^+ \pi^-$ decay amplitude from an unbinned 
maximum-likelihood fit to
the Dalitz plot distribution of a high-purity ($\sim 98\%$) 
$D^0$ sample from 390328 $D^{*+}\to\Dz\pi^+$ decays 
reconstructed in 270 \invfb of data, shown in Fig.~\ref{Fig8}. 

The decay amplitude is expressed as a coherent sum of 
two-body resonant terms and a uniform non-resonant contribution.
For $r=\rho(770)$ and $\rho(1450)$ we use the functional form 
suggested in Ref.~\cite{ref:gounarissakurai}, while the remaining resonances 
are parameterized by a spin-dependent relativistic Breit-Wigner distribution. 
The model consists of 13 resonances leading to 16 two-body decay 
amplitudes and phases 
(see Table~\ref{tab:BWfit-res}), plus the non-resonant contribution, 
and accounts
for efficiency variations across the Dalitz plane and the small 
background contribution.
All the  resonances considered in this model are well established except 
for the two 
scalar $\pi\pi$  resonances, $\sigma$ and $\sigma'$, whose
masses and widths are obtained from our sample~\cite{ref:comment_sigma}. 
Their addition to the model is motivated by an improvement in the 
description of the data.

The possible absence of the
$\sigma$ and $\sigma'$ resonances is considered in the evaluation of
the systematic errors.
In this respect, the K-matrix formalism~\cite{ref:Kmatrix} provides 
a direct way of imposing the unitarity constraint that is not 
guaranteed in the case of the Breit-Wigner parametrization and is suited to the study 
of broad and overlapping resonances in multi-channel decays.
We use the K-matrix method to parameterize
the $\pi\pi$ S-wave states, avoiding the need to introduce the two
$\sigma$ scalars. A description of this alternative parametrization can be 
found in Ref.~\cite{ref:babar_dalitzeps05}.

\begin{table}[!htbp]
\begin{center}
\begin{tabular}{l|c|c|c}
\hline
\\[-0.15in]
    Component  &  $Re\{a_r e^{i\phi_r}\}$ &  $Im\{a_r e^{i\phi_r}\}$ 
& $f_r$ (\%) \\ [0.01in]
\hline \hline
$K^{*}(892)^-$        &  $-1.223\pm0.011$  &  $1.3461\pm0.0096$  &  58.1 \\  
$K^{*}_0(1430)^-$     &    $-1.698\pm0.022$   &   $-0.576\pm0.024$  & 6.7 \\ 
$K^{*}_2(1430)^-$     &    $-0.834\pm0.021$    &  $0.931\pm0.022$  &  6.3 \\  
$K^{*}(1410)^-$       &   $-0.248\pm0.038$  &   $-0.108\pm0.031$    & 0.1 \\ 
$K^{*}(1680)^-$       &    $-1.285\pm0.014$     &   $0.205\pm0.013$    &  0.6\\
\hline
$K^{*}(892)^+$        &   $0.0997\pm0.0036$   & $-0.1271\pm0.0034$   & 0.5 \\
$K^{*}_0(1430)^+$     &    $-0.027\pm0.016$    &    $-0.076\pm0.017$   & 0.0 \\
$K^{*}_2(1430)^+$     &  $0.019\pm0.017$   &  $0.177\pm0.018$   &  0.1 \\ 
\hline
$\rho(770)$           &     $1$     &      $0$                    &   21.6 \\ 
$\omega(782)$         &  $-0.02194\pm0.00099$  & $0.03942\pm0.00066$   & 0.7 \\
$f_2(1270) $          &    $-0.699\pm0.018$   &   $0.387\pm0.018$   & 2.1 \\
$\rho(1450)$          &    $0.253\pm0.038$    &   $0.036\pm0.055$    &   0.1 \\
\hline 
Non-res          &    $-0.99\pm0.19$   &     $3.82\pm0.13$    &    8.5 \\ 
$f_0(980) $           &   $0.4465\pm0.0057$   &   $0.2572\pm0.0081$   &  6.4 \\
$f_0(1370) $          &   $0.95\pm0.11$   &    $-1.619\pm0.011$      &  2.0 \\ 
$\sigma$              &    $1.28\pm0.02$   &  $0.273\pm0.024$    &   7.6 \\ 
$\sigma '$            &  $0.290\pm0.010$   &   $-0.0655\pm0.0098$    &   0.9 \\
\hline
\end{tabular}
\end{center}
\caption{Complex amplitudes $a_r e^{i\phi_r}$ and fit fractions of the 
different components ($K_S\pi^-$, 
$K_S\pi^+$, and $\pi^+\pi^-$ resonances) obtained from the fit of the 
$D^0 \to K_S\pi^+\pi^-$ Dalitz 
distribution from $D^{*+} \to D^0 \pi^+$ events. Errors are statistical 
only.}
\label{tab:BWfit-res}
\end{table}

\section{Dalitz plot analysis of \Dskkp}

\begin{table*}[htb]
\centering
\begin{tabular}{cccc}
\hline
Decay Mode                        & Decay fraction(\%)             & Amplitude              & Phase(radians) \\
\hline
\hline
$\bar K^*(892)^0 K^+$             & $48.7 \pm 0.2 \pm 1.6$    & 1.(Fixed)                   & 0.(Fixed)         \\
$\phi(1020) \pi^+$                & $37.9 \pm 0.2 \pm 1.8$    & $\quad 1.081 \pm 0.006 \pm 0.049$ & $\quad 2.56 \pm 0.02 \pm 0.38$   \\
$f_0(980) \pi^+$                  & $35   \pm 1   \pm 14$     & $4.6  \pm 0.1 \pm 1.6$      & $ -1.04 \pm 0.04 \pm 0.48$\\
$\bar K^*_0(1430)^0 K^+$          & $2.0  \pm 0.2 \pm 3.3$    & $1.07  \pm 0.06 \pm 0.73$   & $-1.37 \pm 0.05 \pm 0.81$  \\
$f_0(1710) \pi^+$                 & $2.0  \pm 0.1 \pm 1.0$    & $0.83  \pm 0.02 \pm 0.18$   & $-2.11 \pm 0.05 \pm 0.42$ \\
$f_0(1370) \pi^+$                 & $6.3  \pm 0.6 \pm 4.8$    & $1.74  \pm 0.09 \pm 1.05$   & $-2.6 \pm 0.1 \pm 1.1$\\
$\bar K^*_0(1430)^0 K^+$          & $0.17  \pm 0.05 \pm 0.30$ & $0.43  \pm 0.05 \pm 0.34$   & $-2.5 \pm 0.1 \pm 0.3$  \\
$f_2(1270) \pi^+$                 & $0.18  \pm 0.03 \pm 0.40$ & $0.40  \pm 0.04 \pm 0.35$   & $0.3 \pm 0.2 \pm 0.5$\\
\hline
\end{tabular}
\caption{
The results obtained from the $D_s^+ \to K^+ K^- \pi^+$ Dalitz plot 
fit, listing fit-fractions, amplitudes and phases. The errors are statistical 
and systematic, 
respectively. The results are preliminary.}
\label{tab:res}
\end{table*}

We study the decay $D^+_s \to K^+ K^- \pi^+$ using  
a data sample of 240 \invfb. We focus particularly on the 
measurement of the relative decay rates 
${{\cal{B}}(D^+_s \to \phi \pi^+)} \over {{\cal{B}}(D^+_s \to K^+ K^- \pi^+)}$ 
and 
${{\cal{B}}(D^+_s \to  \bar K^{*0}(892) K^+)} \over {{\cal{B}}(D^+_s \to K^+ K^- \pi^+)}$.
The decay $D^+_s \to \phi(1020) \pi^+$
is frequently used as the $D^+_s$ reference decay mode.
The improvement in the measurements of these ratios is therefore important. A previous 
Dalitz plot analysis of this decay used $\sim 700$ signal events~\cite{e687}. We 
perform the present analysis using a number of signal events 
more than two orders of magnitude larger.

We reconstruct the decay by fitting the three charged tracks in the event 
to a common vertex, requiring the $\chi^2$ probability to be  
greater than 0.1\%. We cleanly remove 
a small background from the decay $D^{*+} \to D^0_{K^+ K^-}\pi^+$ 
by requiring $m_{K^+ K^-} < 1.85~\gevcc$. In Fig.~\ref{Fig9} we show 
the invariant mass distribution of the reconstructed $D^+_s$ 
candidate in the decay $\Dskkp$. 
For the Dalitz plot analysis, we use events in the 
$\pm 2 \sigma$ mass window of the reconstructed $D^+_s$ candidate.   
We parametrize the incoherent background shape empirically using the 
events in the sidebands.
In the signal region, we find 100850 signal events with a purity 
of about 95\%.

The Dalitz plot for the $D^+_s \to K^+ K^- \pi^+$ events is shown 
in Fig.~\ref{Fig10}. In the $K^+ K^-$ threshold region, a strong 
$\phi(1020)$ signal can be observed, 
together with a rather broad structure indicating the presence of 
the $f_0(980)$ and $a_0(980)$ \textit{S-}wave resonances.
A strong $K^{*0}(890)$ signal can also be seen.
We perform an unbinned maximum likelihood fit 
to determine the relative amplitudes and phases 
of intermediate resonant and non-resonant states.
The complex amplitude coefficient for each of the contributing 
states is measured with respect to $\overline{K}{}^{*0} K^+$. 
We summarize the fit results in Table~\ref{tab:res} showing  
fit-fractions, amplitudes, and phases of the contributing resonances. 
The projections of the Dalitz plot variables in data and the ones from the fit
results are shown in Fig.~\ref{Fig10}. 
Further tests on the fit quality can be estimated using 
$Y^0_L$ angular moments. These moments are shown for the 
$K^+K^-$ and $K^-\pi^+$ channels in Fig.~\ref{Fig11}.
The agreement between the data and fit is excellent. We find a 
rather large contribution from the $f_0(980)\pi^+$, but 
with a large systematic uncertainty due primarily to a poor knowledge 
of the shape parameters of $f_0(980)$ and higher $f_0$ states. 

From the fit-fraction values reported in Table~\ref{tab:res}, we make 
the following preliminary measurements:\\

\noindent
${{\cal{B}}(D^+_s \to \phi \pi^+)} \over {{\cal{B}}(D^+_s \to K^+ K^- \pi^+)}$
$=$ 0.379 $\pm$ 0.002 (stat) $\pm$ 0.018 (syst), \\ 

\noindent
${{\cal{B}}(D^+_s \to \bar K^{*0}(892) K^+)} \over {{\cal{B}}(D^+_s \to K^+ K^- \pi^+)}$
$=$ 0.487 $\pm$ 0.002 (stat) $\pm$ 0.016 (syst). \\

\section{Conclusions}
we have studied the amplitudes of the decays \Dzkkpz, \Dzpppz, \Dzkspp, 
and \Dskkp. Using \Dzkkpz\ Dalitz plot analysis, we 
measure the strong phase difference between the \Dzb\ and \Dz\ decays 
to $K^*(892)^{+} K^-$ and their amplitude ratio, 
which will be useful in the measurement of the CKM phase $\gamma$.  
We observe contributions from the $K\pi$ and $K^-K^+$ scalar and 
vector amplitudes, and analyze their angular moments. We find no evidence for 
charged $\kappa$, nor for higher spin states. We also 
perform a partial-wave analysis of the $K^-K^+$ system in a limited mass 
range. We measure the magnitudes and phases of the components of the 
$\Dz\to\ppp$ decay amplitude, which we use in constraining the 
CKM phase $\gamma$ using $\decaychain$. We measure the amplitudes of 
the neutral $D$-meson decays to the $K^0_s\pi^-\pi^+$ final state and use 
the results as input in the measurement of $\gamma$ using the decay 
$B^\mp\rightarrow D^{(*)}_{K^0_s\pi^-\pi^+}K^\mp$. Finally we parametrize  
the amplitudes of the $D_s^+ \to K^+ K^- \pi^+$ Dalitz plot and perform  
precision measurements of the relative decay rates 
${{\cal{B}}(D^+_s \to \phi \pi^+)} \over {{\cal{B}}(D^+_s \to K^+ K^- \pi^+)}$ 
and 
${{\cal{B}}(D^+_s \to  \bar K^{*0}(892) K^+)} \over {{\cal{B}}(D^+_s \to K^+ K^- \pi^+)}$.
\section{Acknowledgements}
We are grateful for the excellent luminosity and machine conditions
provided by our PEP-II colleagues, and for the substantial dedicated effort 
from the computing organizations that support \babar. 
This work is supported by the United States Department of Energy
and National Science Foundation.


\begin{thebibliography}{99}
\bibitem{note1}
Reference to the charge-conjugate decay is implied throughout. The initial 
state referred to is \Dz, not \Dzb.
\bibitem{ref:D} 
We use the symbol $D$ to indicate any linear combination of
a $\Dz$ and a $\Dzb$ meson state.
\bibitem{ref:comment_sigma} 
The $\sigma$ and $\sigma'$ masses and widths are determined 
from the data. We find (in~\mevcc) 
$M_{\sigma}=490\pm6$, $\Gamma_{\sigma}=406\pm11$, 
$M_{\sigma'}=1024\pm4$, and $\Gamma_{\sigma'}=89\pm7$. 
Errors are statistical.
\end{thebibliography}
\end{document}